\documentclass{aastex}
\usepackage[onecolumn]{emulateapj5}

\newcommand{\civ}{\ion{C}{4}}

\newcommand{\hi}{\ion{H}{1}}

\newcommand{\kms}{km\,s$^{-1}$}
\newcommand{\kmsm}{km\,s$^{-1}$\,Mpc$^{-1}$}

\newcommand{\mgii}{\ion{Mg}{2}}

\newcommand{\oii}{[\ion{O}{2}]}

\newcommand{\cai}{\ion{Ca}{1}}
\newcommand{\caii}{\ion{Ca}{2}}
\newcommand{\CaH}{\caii\,$\lambda$3969}
\newcommand{\CaK}{\caii\,$\lambda$3934}
\newcommand{\fei}{\ion{Fe}{1}}
\newcommand{\feii}{\ion{Fe}{2}}
\newcommand{\feiii}{\ion{Fe}{3}}

\newcommand{\hei}{\ion{He}{1}}
\newcommand{\HeI}{\ion{He}{1}\,$\lambda$3188}
\newcommand{\HEI}{\ion{He}{1}\,$\lambda$3889}
\newcommand{\heii}{\ion{He}{2}}

\newcommand{\lya}{Ly$\alpha$}

\newcommand{\mgi}{\ion{Mg}{1}}

\newcommand{\MgI}{\mgi\,$\lambda$2852}

\newcommand{\TIii}{Ti\,{\sc ii}}

\shorttitle{CA II LOBAL QUASAR}
\shortauthors{HALL ET AL.}

\begin{document}

\journalinfo{Accepted to ApJ Apr. 18, 2003, for the Aug. 10, 2003 issue}
\submitted{}
\title{VLT+UVES Spectroscopy of the \ion{Ca}{2} LoBAL Quasar 
SDSS J030000.56+004828.0\altaffilmark{1}}
\altaffiltext{1}{Based on observations from ESO 
Director's Discretionary Time program 267.A-5698.}

\author{
Patrick B. Hall\altaffilmark{2,3},
Damien Hutsem\'ekers\altaffilmark{4,5},
Scott F. Anderson\altaffilmark{6},
J. Brinkmann\altaffilmark{7},
Xiaohui Fan\altaffilmark{8,9},
Donald P. Schneider\altaffilmark{10},
Donald G. York\altaffilmark{11,12}
}
\altaffiltext{2}{Princeton University Observatory, Princeton, NJ 08544}
\altaffiltext{3}{Departamento de Astronom\'{\i}a y Astrof\'{\i}sica, Facultad
de F\'{\i}sica, Pontificia Universidad Cat\'{o}lica de Chile, 
Casilla 306, Santiago 22, Chile}
\altaffiltext{4}{Research Associate FNRS, University of Li\`ege,
     All\'ee du 6 ao\^ut 17, Bat. 5c, 4000 Li\`ege, Belgium}
\altaffiltext{5}{European Southern Observatory, Casilla 19001, Santiago, Chile}
\altaffiltext{6}{University of Washington, Department of Astronomy, Box 351580, Seattle, WA 98195}
\altaffiltext{7}{Apache Point Observatory, P.O. Box 59, Sunspot, NM 88349-0059}
\altaffiltext{8}{Institute for Advanced Study, Olden Lane, Princeton, NJ 08540}
\altaffiltext{9}{Current affiliation: Steward Observatory, The University of Arizona, Tucson, Arizona 85721}
\altaffiltext{10}{Department of Astronomy and Astrophysics, The Pennsylvania State University, University Park, PA 16802}
\altaffiltext{11}{The University of Chicago, Department of Astronomy and Astrophysics, 5640 S. Ellis Ave., Chicago, IL 60637}
\altaffiltext{12}{The University of Chicago, Enrico Fermi Institute, 5640 S. Ellis Ave., Chicago, IL 60637}

\begin{abstract}
We study high-resolution spectra of the `overlapping-trough' low-ionization 
broad absorption line (LoBAL) quasar SDSS J030000.56+004828.0.
The \caii, \mgii\ and \mgi\ column densities in this object 
are the largest reported to date for any BAL outflow.
The broad \caii\ absorption 
is mildly blended, but the blending can be
disentangled to measure the \caii\ column density,
which is large enough that the outflow must include 
a strong hydrogen ionization front.  
The outflow begins at a blueshift of $\sim$1650\,\kms\ from the systemic 
redshift.  The lowest velocity BAL region produces strong \caii\ absorption
but does not produce significant excited \feii\ absorption, 
while the higher velocity excited \feii\ absorption region
produces very little \caii\ absorption.
We have found that only a disk wind outflow can explain this segregation.
Whether the outflow is smooth or clumpy, we conclude that
the \caii\ BAL region has a density high enough to 
populate excited levels of \feii, but a temperature 
low enough to prevent them from being significantly populated.
This requirement means the \caii\ BAL region has $T \lesssim 1100$~K,
and perhaps even $T \lesssim 550$~K.
This quasar also has an associated absorption line system (AAL) that exhibits
partial covering, and therefore is likely located near the central engine. 
Its association with the BAL outflow is unclear.
Blending of the AAL with the BAL trough shows that
the spatial region covered by the BAL outflow can vary
over velocity differences of $\sim$1700~\kms.
\end{abstract}


\section{Introduction}  \label{INTRO}

One of the goals of the Sloan Digital Sky Survey \markcite{yor00}(SDSS; {York} {et~al.} 2000) 
is to obtain spectra for $\sim$10$^5$ quasars to $i=19.1$ ($i=20.2$ for $z>3$
candidates), in addition to the $\sim10^6$ galaxies that comprise the bulk
of the spectroscopic targets \markcite{sdss89}({Blanton} {et~al.} 2003).  From astrometrically calibrated
drift-scanned imaging data \markcite{gun98,sdss153}({Gunn} {et~al.} 1998; {Pier} {et~al.} 2003) on the SDSS $ugriz$ 
AB asinh magnitude system \markcite{fuk96,sdss26,sdss82,sdss85,sdss105}({Fukugita} {et~al.} 1996; {Lupton}, {Gunn}, \& {Szalay} 1999; {Hogg} {et~al.} 2001; {Stoughton} {et~al.} 2002; {Smith} {et~al.} 2002),
quasar candidates are selected primarily using color criteria designed to
target objects whose broad-band colors are different from those of normal stars
and galaxies \markcite{sdssqtarget}({Richards} {et~al.} 2002).
Due to these inclusive criteria, the selection of candidates using $i$ band
magnitudes rather than blue magnitudes
(which are more affected by absorption and reddening),
and its area and depth, the SDSS is effective at finding 
unusual quasars \markcite{fan99b}(e.g., {Fan} {et~al.} 1999).
Many of these objects are unusual broad absorption line (BAL) quasars.

BAL quasars show absorption from gas with typical
outflow velocities of 0.1$c$ \markcite{wea91}({Weymann} {et~al.} 1991)  
and mass loss rates probably
comparable to the accretion
rates required to power quasars ($\sim1$~$M_{\odot}$\,yr$^{-1}$).  
Therefore an understanding of BAL outflows is required for an understanding of
quasars as a whole.
Usually BAL troughs are only seen in high ionization species, but about 15\% of
BAL quasars also show absorption from low ionization species, including
absorption from excited fine-structure levels or terms of \feii\ and \feiii\ in
rare cases.  
In \markcite{sdss123}{Hall} {et~al.} (2002, hereafter H02)
we presented several examples of `overlapping-trough'
BAL quasars whose flux shortward of \mgii\ is almost completely absorbed by
troughs of excited \feii\ (which is of course also accompanied by ground-state
\feii\ absorption).
In this paper we present a high-resolution spectrum of one such object,
SDSS J030000.56+004828.0 (hereafter simply SDSS J0300+0048) at $z=0.89$, 
and use it to constrain the column densities and possible
structure of its impressive outflow.
SDSS J0300+0048 is
only the fourth quasar known to have a \caii\ BAL trough (\S\,5.2.1 of 
H02),\footnote{\markcite{rvs02}{Rupke}, {Veilleux}, \& {Sanders} (2002) 
find that ultraluminous infrared galaxies also can
have \caii\ outflows with large rest-frame equivalent widths (REWs).
It is unclear how similar these narrower outflows are to \caii\ BALs.
The continuum around \caii\ in most of their objects is dominated by galaxy
light; the addition of a quasar continuum might greatly decrease the REWs.
In any case, the outflows in SDSS J0300+0048 and QSO J2359$-$1241
have much larger blueshifts than
the outflows studied by \markcite{rvs02}{Rupke} {et~al.} (2002).}
and only the second to be studied at high resolution, after QSO J2359$-$1241
\markcite{aea01}({Arav} {et~al.} 2001). 
Both quasars have ``detached" 
absorption troughs that do not start at the systemic redshift:
those in QSO J2359$-$1241 set in at a blueshift of $\sim$750\,\kms, 
and those in SDSS J0300+0048 at a blueshift of $\sim$1700\,\kms. 
SDSS J0300+0048 also
has a strong, narrow associated absorption system at the systemic redshift,
with \mgii\ REW=1.882\,\AA.
Whether this system is intrinsic to 
the quasar is one issue we consider in this paper.

We begin with a discussion of the most common method of analyzing intrinsic
absorption lines, then apply this method to SDSS J0300+0048,
and finally discuss our results.

\section{Analyzing Intrinsic Absorption Lines} \label{ANALYSIS} 

Suppose we observe unblended doublet absorption from a BAL region covering a
fraction $C$ of the quasar's light with an optical depth $\tau$.
The resulting normalized residual intensities 
in the two lines of the doublet as a function of velocity $v$ are: 
\begin{equation}                 \label{e_I1}
I_1(v)=1-C_v(1-e^{-\tau_v})
\end{equation}                 
\begin{equation}                 \label{e_I2}
I_2(v)=1-C_v(1-e^{-R\tau_v})
\end{equation}                 
where $R=g_2f_2\lambda_2/g_1f_1\lambda_1$ is the ratio of the optical depth
of line 2 to that of line 1 (the $g_i$ are the statistical weights, the $f_i$
the oscillator strengths, and the $\lambda_i$ the wavelengths of the lines).
Our convention is that $I_1$ refers to the stronger line of the doublet
and $\tau_v$ refers to the optical depth in that line.
For any doublet with $R=0.5$, such as \mgii\ and \caii,
these equations yield the following solutions \markcite{hbjb97}({Hamann} {et~al.} 1997)
and associated uncertainties for
$C_v$ and $\tau_v$:
\begin{equation}                 \label{e_C}
C_v={{1+I_2^2-2I_2}\over{1+I_1-2I_2}};
~~\sigma_{C_v}=\frac{
\sqrt{\sigma_{I_1}^2(I_2-1)^4
+\sigma_{I_2}^2[2(I_2-1)(2+I_1-3I_2)]^2}}{(1+I_1-2I_2)^2}
\end{equation}
\begin{equation}                 \label{e_tau}
\tau_v=-\ln \left( {{I_1-1+C_v}\over{C_v}} \right)
=-2\ln \left( {{I_2-I_1}\over{1-I_2}} \right);
~~\sigma_{\tau_v}=\frac{2e^{\tau_v}(I_2-I_1)}{(I_2-1)^2}
\sqrt{\sigma_{I_1}^2+\sigma_{I_2}^2\left(\frac{I_1-1}{I_2-1}\right)^2}
\end{equation}
where it is understood that $I_1$ and $I_2$ are functions of velocity.
The solution for $C_v$, and thus for $\tau_v$ as well, is only physical 
($0 \leq C_v \leq 1$) when
\begin{equation}		\label{e_valid}
0\leq I_1\leq 1  ~~~{\rm and}~~~ 0\leq I_2\leq 1 ~~~{\rm and}~~~ 
I_2^2\leq I_1 ~~~{\rm and}~~~ I_2\geq I_1
\end{equation}
The last of the above conditions could be violated in the case of saturated
absorption partially filled in by optically thin emission from the same
doublet transition responsible for the absorption.
The next to last could be invalidated by the presence in the spectrum of
significant host galaxy or scattered quasar light.

The above equations can also break down if there are different covering factors
for different emission regions (e.g., the continuum source, the broad emission
line (BEL) region, and the \feii\ emission region; see \S\,\ref{VRB} and
\markcite{gecc99}{Ganguly} {et~al.} 1999) or if the emissivities of the
covered and uncovered regions are different functions of velocity 
(as pointed out by \markcite{ss99}{Srianand} \& {Shankaranarayanan} 1999 
and observed by, e.g., \markcite{gea03}{Gabel} {et~al.} 2003).
Moreover, the assumption that a fraction $C$ of the quasar's light is occulted
at a single optical depth $\tau$ and that the remaining light is unobscured
is likely to itself be an approximation 
\markcite{dka02}({de Kool}, {Korista}, \&  {Arav} 2002b), and would be 
inappropriate if there are cases where the resonance scattering interpretation
of BAL quasars is correct \markcite{bra02}({Branch} {et~al.} 2002).
Nevertheless, the above equations are useful for studying BAL troughs, even if 
only to verify that one or more of the above complications is occurring.

If the absorption is broad enough that the two lines of the doublet
are blended over a velocity width $v_b$ that is less than their velocity
separation $\Delta v$,\footnote{See 
\markcite{jbs83}{Junkkarinen}, {Burbidge} \& {Smith} (1983)
for an approach to the $v_b > \Delta v$ case.}
then there are three distinct velocity regimes to consider,
which we label A to C from high velocity to low.  In A and C one of the 
two lines produces blended absorption, while in B neither line is blended;
see \S\,\ref{OVER0300} for a concrete example.
Using positive $v$ to denote outflows (absorption at shorter wavelengths) and
assuming the stronger line (subscript 1) is the shorter wavelength one yields:
\begin{eqnarray}                 \label{e_Ilampp}
A:~~~~~~~~~~~~~~~~~v_{max}>v>v_{max}-\Delta v+v_b:~~~~~~
I_1=1-C_v(1-e^{-\tau_v})~~~~~~~~~~~~~~~~~~~~~~~~~~~~~~~~~~~~~~~~\\
I_2=[1-C_v(1-e^{-R\tau_v})] \otimes [1-C_{v-\Delta v}(1-e^{-\tau_{v-\Delta v}})]\\
B:~~~v_{max}-\Delta v+v_b>v>v_{max}-\Delta v:~~~~~~~~~~~~
I_1=1-C_v(1-e^{-\tau_v})~~~~~~~~~~~~~~~~~~~~~~~~~~~~~~~~~~~~~~~~\\
I_2=1-C_v(1-e^{-R\tau_v})~~~~~~~~~~~~~~~~~~~~~~~~~~~~~~~~~~~~~~~\\
C:~~~~~~~~~v_{max}-\Delta v>v>v_{max}-\Delta v-v_b:~~~~~~
I_1=[1-C_v(1-e^{-\tau_v})] \otimes [1-C_{v+\Delta v}(1-e^{-R\tau_{v+\Delta v}})]\\
I_2=1-C_v(1-e^{-R\tau_v})~~~~~~~~~~~~~~~~~~~~~~~~~~~~~~~~~~~~~~~
\end{eqnarray}
where $v_{max}$ is the maximum outflow velocity and
$C_{v\pm\Delta v}$ and $\tau_{v\pm\Delta v}$
are the covering factors and optical depths at velocities $v\pm\Delta v$.
If blending does not occur ($v_b=0$), regimes $B$ and $C$ vanish and
regime $A$ reduces to Eqs. \ref{e_I1}-\ref{e_I2} because $C_{v\pm\Delta v}$
and $\tau_{v\pm\Delta v}$ become undefined --- both $C$ and $\tau$
will be defined only over a velocity range less than $\Delta v$.
Otherwise, the convolution symbols indicate that the observed spectrum is
affected by absorption from both lines in a possibly complicated manner.
Nonetheless, because $C_{v\pm\Delta v}$ and $\tau_{v\pm\Delta v}$ are just the
values of $C_v$ and $\tau_v$ in regime B, which can be solved for, it may be
possible to estimate the $C_v$ and $\tau_v$ in regimes A and C.
For example, if $C_v$ is a constant and a fixed emission region is covered
at all velocities, the convolutions are simple multiplications.
This is not the case if $C_v$ varies or if different emission regions are
covered as a function of velocity, even if $C_v$ is numerically constant.

\section{VLT + UVES Observations of SDSS J0300+0048} \label{UVES} 

Observations of several SDSS BAL quasars were obtained on UT 10-12 Aug 2001
using the ESO Very Large Telescope (VLT) Unit 2 (Kueyen) 
and Ultra-Violet Echelle Spectrograph (UVES).
This section provides observational details for all objects; science results
for SDSS J2215$-$0045 and SDSS J1453+0029 can be found in 
\markcite{sdss199}{Hall} \& {Hutsem\'ekers} (2003),
and for SDSS J0011$+$0055 in Hutsem\'ekers {et~al.} (2003, in preparation).
A 1\arcsec\ slit was used, yielding a resolution $R \simeq 40,000$ (7.5 \kms)
at all wavelengths.  A depolarizer was also used for all observations.
Each exposure was reduced individually with optimum extraction \markcite{hor86}({Horne} 1986),
including simultaneous background and sky subtraction.
For SDSS~J0300+0048, wavelengths 3030$-$3880\,\AA\ and 4760$-$6840\,\AA\
were observed simultaneously, in two exposures of 4500 and 2992 seconds,
and wavelengths 3730$-$4990\,\AA\ and 6600$-$10600\,\AA\ simultaneously
in a single exposure of 4500 seconds.
Because the 3030$-$3880\,\AA\ 
setting had a very low signal-to-noise ratio
and yielded different results for the two exposures, it is not used in this
paper.  The extraction for the 3730$-$4990\,\AA\ 
setting
was determined manually in places where the flux was zero over several orders.
Telluric absorption lines were removed for the red settings using observations
of telluric standard stars, shifted in velocity according to the different times
of the observations and scaled in intensity according to the airmass difference.
Whenever two or three exposures of each setting were available, their extracted
spectra were averaged with rejection of cosmic rays and known CCD artifacts.
Finally, all settings were rebinned to a constant wavelength interval of
0.015\,\AA\ on a vacuum heliocentric scale, scaled in intensity to match
each other in their overlap regions, and merged into a single spectrum.

\section{Analyzing SDSS J0300+0048} \label{S0300} 

Figure \ref{f_0300sb2} shows a portion of the SDSS spectrum of SDSS J0300+0048,
intended to help put in context the small regions of high-resolution UVES
spectra we will present in this paper.
From the central subsystem of the associated absorber
seen in \caii\,$\lambda$3969 and \mgi\ in the UVES spectrum,
we adopt a systemic redshift of $z=0.891850\pm0.000005$,
in excellent agreement with the value $z=0.89191\pm0.00005$ measured 
from the narrow \mgii\ absorption in the SDSS spectrum (H02). 
There is weak \oii\ emission at $z=0.8908\pm0.0007$, 
which is a blueshift of 166$\pm$111~\kms\ 
(using 3728.48\,\AA\ for the unresolved doublet's vacuum rest wavelength).

\subsection{What is the Continuum?}	\label{WHAT}

To analyze the absorption in SDSS J0300+0048 in a rigorously correct manner,
we must know or be able to model its intrinsic, unabsorbed continuum at the 
wavelengths where absorption is present.  Constructing such a continuum for
the transitions of interest in SDSS J0300+0048 (\mgii, \mgi\ and \caii)
requires the matching of complex blends of iron emission and absorption.

Figure \ref{f_0300fe2fit}
shows the SDSS rest frame spectrum of SDSS J0300+0048 
(solid line).  The emission features longward of \mgii\ are plausibly
identifiable as blends of \feii\ and \fei, as discussed in \S\,5.2 of H02,
though we now recognize that \TIii\ emission is also likely to contribute
\markcite{kea95}({Kwan} {et~al.} 1995).
The dashed line shows one possible model for the unabsorbed continuum of
SDSS J0300+0048,
consisting of a flat-in-$F_{\lambda}$ power-law plus a scaled and smoothed
iron emission template.  This template
was constructed by fitting and subtracting a power-law to the spectrum
of the bright, narrow-lined, strong \feii-emitting quasar 
SDSS J092332.33+574557.4 (D. P. Schneider {et~al.}, in preparation),
using continuum windows at 1680-1700\,\AA\ and 3675-3710\,\AA.
Because the template is intended for illustrative purposes only, no attempt was
made to exclude emission from \mgii\ or other non-iron transitions or to
optimize the slope of the power-law continuum component of the model spectrum.
This model spectrum is a good match to the emission feature just shortward
of 3000\,\AA, the blue side of which may be affected by absorption,
and to the wavelength extent of the weak feature at 3400-3600\,\AA.  However,
this latter feature is much stronger in SDSS J0300+0048 than in the model,
SDSS J0300+0048 has emission features at 3025\,\AA\ and 3250\,\AA\ not seen
in the model (the latter could be \TIii), and
there are two features around 3100\,\AA\ in the template that are not seen in
SDSS J0300+0048.  Given these discrepancies between the model spectrum and the
observed spectrum, we cannot use the model spectrum to predict the intrinsic
continuum of SDSS J0300+0048 at $\lambda$$<$2900\,\AA\ with any confidence.
In the future it may be possible to find another strong \feii-emitting quasar
whose spectrum better matches that of SDSS J0300+0048 at $\lambda$$>$3000\,\AA,
but this is the best match we have found to date in the SDSS or in the
literature (e.g., \markcite{gcc96}{Graham}, {Clowes}, \& {Campusano} 1996).  Note that the
Fe emission templates of \markcite{bg92}{Boroson} \& {Green} (1992) and \markcite{vw01}{Vestergaard} \& {Wilkes} (2001) only cover
4250-7000\,\AA\ and 1250-3090\,\AA, respectively.

Alternatively,
if we could match the $\lambda$$>$3000\,\AA\ spectrum of SDSS J0300+0048 using
a theoretical model, we could use that model to predict its unabsorbed
continuum at $\lambda$$<$3000\,\AA.
However, published theoretical \feii\ emission models either do not extend
to $\lambda$$>$3200\,\AA\ \markcite{ver99}({Verner} {et~al.} 1999) or do not include sufficient
transitions at $\lambda$$>$3200\,\AA\ to match
observations \markcite{sp03}({Sigut} \& {Pradhan} 2003),
and the situation for other possibly important ions such as \fei\ and
\TIii\ is no better.

Thus at present we have no reliable model for the unabsorbed continuum of
SDSS J0300+0048 at the wavelengths of \mgii, \mgi\ or \caii.  For \caii,
we can construct a reasonable local continuum.
For \mgii, \mgi\ and other transitions at even shorter wavelengths we must 
keep in mind that a local continuum estimate probably
underestimates the unabsorbed continuum by a factor of $\sim$2,
unless dust reddening produces a turnover in the unabsorbed spectrum.

\subsection{The Associated Absorption} \label{ASSABS}

To remove the narrow associated \CaK\ blended with the broad \CaH,
we must model the associated system.
This procedure yielded interesting results, but ones tangential
to the main paper, and so is discussed in Appendix \ref{APPABS}.

\subsection{The Broad Absorption} \label{BAL0300}

\subsubsection{Overview} \label{OVER0300}

Figure \ref{f_broad5way} shows the normalized spectrum of SDSS J0300+0048 
around the broad absorption troughs of five separate transitions.
The heavier lines denote spectral regions that are not confused with absorption
from different transitions.
Each transition has unique, unconfused velocity ranges, as we now discuss
(from top to bottom in Figure \ref{f_broad5way}).

The \mgii\ trough starts at about 1650~\kms, and is at least 5500~\kms\ wide;
\mgii\ is confused with \feii\,$\lambda$2773 (UV63) at $>$7150~\kms.  An upper
limit to the \mgii\ trough width is impossible to determine, given that the
continuum does not return to its unabsorbed level anywhere shortward of \mgii\ 
(Figure \ref{f_0300sb2}).  However, a reasonable maximum velocity can be
inferred from the \feii\,UV63 absorption, which disappears by $\sim$10850~\kms\ 
for the UV63 multiplet line with the shortest wavelength (2715\,\AA).
That maximum velocity implies a \mgii\ trough at least 9200~\kms\ broad.

The \mgi\ absorption appears to start when the \mgii\ trough
appears to saturate, at $\sim$2000~\kms.
It is unclear just how high a velocity the \mgi\ absorption reaches,
both because \mgi\ is confused with associated \mgii\ at $>$5200~\kms\ and
because the continuum normalization between
\mgi\ and \mgii\ is very uncertain.  The \mgi\ 
might reach only $\sim$4000~\kms, matching the strong \caii\ absorption,
in which case the alternate continuum normalization shown by the tilted dashed
line in the Figure would be more appropriate.
Alternatively, the \mgi\ 
might reach $\sim$6400~\kms, matching the possible highest-velocity \caii, 
or even extend to $>$7150~\kms\ --- like \mgii\ --- in which case
it would overlap with
\mgii.

The broad \caii\ was normalized by a simple linear continuum fit to 
two narrow wavelength regions around 7294\,\AA\ and 7478\,\AA.
To determine the blending in \caii, we use the fact that 
\caii\,$\lambda$3934 is the stronger line of the doublet, which means
we can be sure that the highest velocity detectable absorption
is from it and not from even higher velocity \caii\,$\lambda$3969.
The \caii\,$\lambda$3934 absorption does not appear to reach velocities as
large as the \mgii\ absorption does, but instead disappears by $\sim$5657~\kms. 
Given that the velocity separation of the \caii\ doublet is 2640.1~\kms,
the \caii\,$\lambda$3934 trough is unblended only at 3017-5657~\kms.
Similarly, because the \caii\,$\lambda$3969 absorption only starts at 1697~\kms,
the \caii\,$\lambda$3969 trough is unblended only at 1697-4337~\kms\ 
(discounting the narrow associated \caii\,$\lambda$3934 near 2640~\kms).

Ground-state \feii\ absorption is undoubtably present at all velocities where
\mgii\ is seen, because the density and ionization conditions under which the
two ions exist are very similar.  However, at the lowest velocities in the
outflow, the fine-structure levels of \feii\ are not populated and so excited
\feii\ absorption is not seen at those velocities, as first noted in H02.
The \feii\,$\lambda$2632 absorption 
(arising from an excited fine-structure level in the ground term,
this is the longest wavelength line of the UV1 multiplet)
is slightly contaminated by associated \feii\,$\lambda$2600
(the strongest transition in the UV1 multiplet, this line is from
the lowest fine-structure level in the ground term, not an excited level).
Nonetheless, it is clear that the broad \feii\,$\lambda$2632 absorption begins
at a velocity that matches the highest-velocity component of the main \caii\ 
trough (see \S\ref{CAFE}). 
As mentioned earlier, 
the high-velocity limit of the \feii\ 
absorption is at least 10850~\kms.

All the above are low-ionization species, so at first glance it may seem
surprising that they have different velocity distributions.  However, the
ionization potentials of \mgii\ and \feii\ are higher than that of \hi,
while those of \mgi\ and \caii\ are lower.  This difference can lead to very
different behavior for the latter two ions, as discussed in \S\ref{CODIS}.

\subsubsection{Broad \caii\ Absorption: Velocity Regime B} \label{VRB}

\begin{figure} \epsscale{1.0} 
\plotone{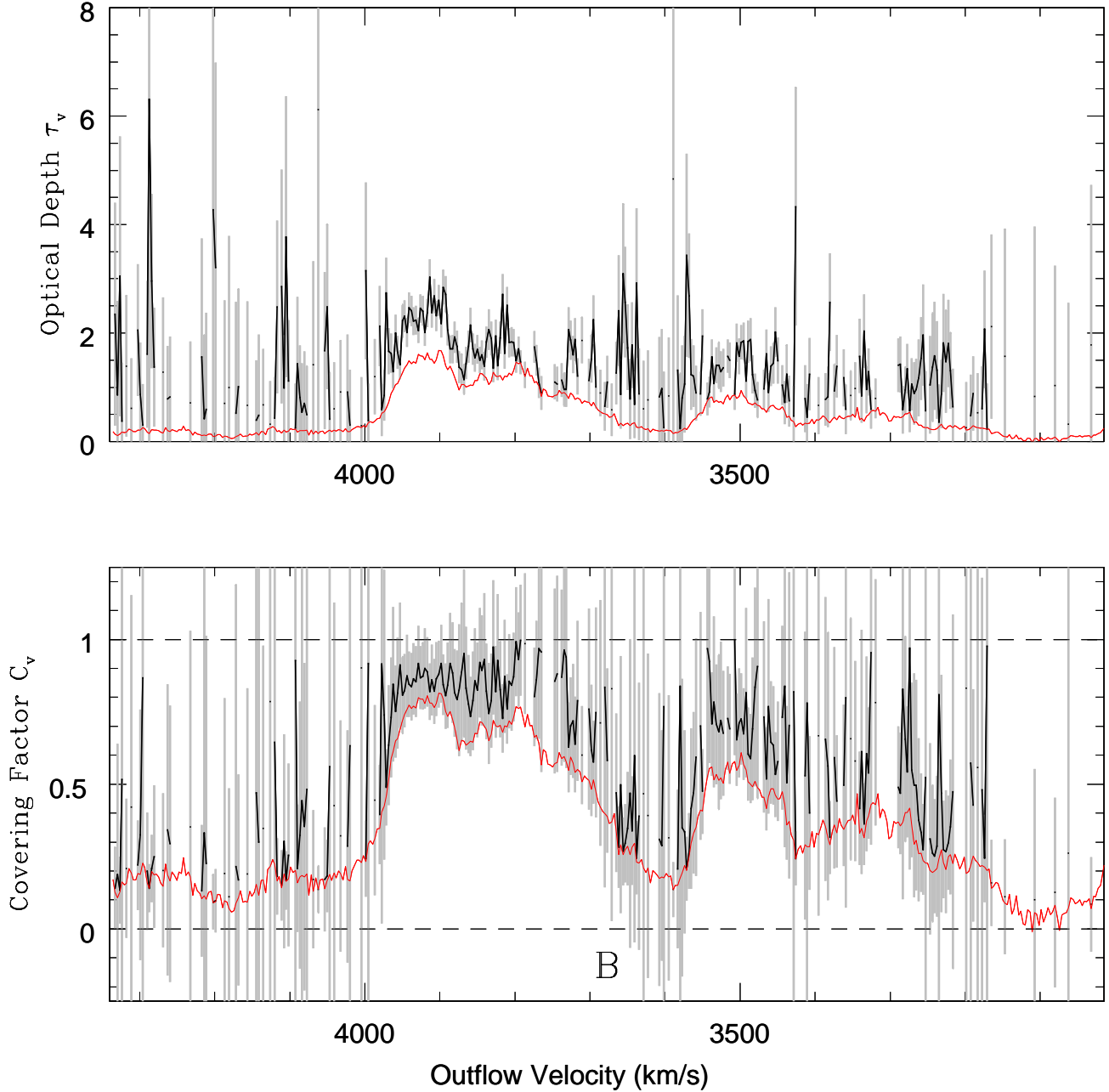}		
\caption[]{ \singlespace 
The thick black line gives the optical depth (top)
and covering factor for \ion{Ca}{2} as a function of velocity
in the (unblended) velocity regime B,
after binning by 5 pixels in the dispersion direction.
Points are plotted only when a physical solution is possible at that velocity.
The grey areas show the $\pm1\sigma$ uncertainty ranges, which can exceed the
range of physical solutions (namely, $0\leq C_v \leq1$, shown by the
dashed lines in the bottom panel).
The thin red line in the top panel shows the minimum optical depth
needed to match the observed \ion{Ca}{2}\,$\lambda$3934 absorption
in the limit of 100\% covering.
The thin red line in the bottom panel shows the minimum covering factor
needed to match the observed \ion{Ca}{2}\,$\lambda$3934 absorption
in the limit of infinite optical depth.
See \S\,\ref{VRB} for discussion.
}\label{f_Cerrtau0300B}
\end{figure}

To begin the quantitative analysis of the broad \caii\ absorption, we used
Eqs. 8-9 to solve 
for $C_v$ and $\tau_v$ in the unblended outflow
at 3017-4337~\kms\ (velocity regime B).
The results are shown in Figure \ref{f_Cerrtau0300B},
where the original pixels have been binned by a factor of 5 in order to
achieve an acceptable signal-to-noise ratio (S/N).
We defer a detailed discussion of this Figure until later in this section.
Note that no points are plotted for velocity bins where the residual
intensities do not allow a physical solution because one or more of the
conditions of Equation~\ref{e_valid} are violated.
This situation occurs for 37\% of the original pixels in velocity regime B.
Inspection of the velocities where the residual intensities did not allow a
physical solution suggested that our continuum normalization was an
underestimate.
Thus we refit the continuum using a third-order Legendre polynomial between
two narrow wavelength regions around 7278\,\AA\ and 7496\,\AA.
This normalization yielded only a slight increase in the number of pixels
where the residual intensities allowed a physical solution, suggesting that
our continuum normalization is not the major source of error in that respect.

One possible source of error is scattered light or host galaxy light that
does not pass through the BAL region and thus fills in the absorption troughs.
Host galaxy light should be a $\lesssim1$\% effect because this quasar is
quite luminous ($M_i=-27.08$ for $H_0=70$\,\kmsm, $\Omega_M=0.3$, and
$\Omega_{\Lambda}=0.7$; D. P. Schneider {et~al.}, in preparation).
The observed minimum residual intensity in the broad \caii\ trough is
6.0$\pm$1.5\% of the continuum, and so that is the maximum percentage that
unabsorbed scattered light or host galaxy light
can contribute to the observed continuum.
To test the effects of a maximal contribution from unabsorbed scattered light,
we subtracted 0.06 from the residual intensity at all velocities, renormalized
by 0.94, and repeated our calculation of $C$ and $\tau$.  This adjustment
slightly decreased the number of pixels with allowed physical solutions,
suggesting that ignoring the contribution of scattered light does not 
greatly affect the calculation of $C_v$ and $\tau_v$.

Another possible source of error is our assumption that the covering factor is
the same for the featureless continuum source and \feii\ emission regions.
Doublet absorption does not provide enough information to solve for two covering
factors
independently.  However,
\S\,4.2 of \markcite{gecc99}{Ganguly} {et~al.} (1999) discusses how joint constraints on $C_c$ and $C_{Fe}$
can still be made, given the ratio $W$ of the \ion{Fe}{2} flux to the continuum
source flux. 
For a lower limit of $W=0.07$ in SDSS J0300+0048, the situation lies in between
that of a single covering factor and the scattered light case
considered previously, so assuming a single covering factor is reasonable.  
In the more realistic $W=0.27$ case, the covering factors have similar, large
values (maxima of $C_c>0.92$ and $C_{Fe}>0.72$) and assuming a single covering
factor again appears reasonable.\footnote{We cannot rule out an \ion{Fe}{2}
emission region that remains uncovered until the continuum emission region is
fully covered, and thus has $C_{Fe}=0$ at most velocities. However,
refitting $C_v$ and $\tau_v$ for such a scenario yields a large increase
in the number of pixels that do not correspond to a physical solution.
}

Neither the details of the assumed continuum, the neglect of scattered light, or
the assumption of a single covering factor appear responsible for the relatively
large fraction of pixels (37\%) for which the residual intensities in the
\caii\ doublet lines do not allow a physical solution for $C_v$ and $\tau_v$.
The explanation for these pixels must lie primarily in the random noise.  This
explanation is consistent with Figure \ref{f_Cerrtau0300B}, which shows that
the bulk of the pixels without physical solutions lie in the regions of weak
absorption, where the noise is comparable to the strength of the absorption.
(The average S/N per pixel in the \caii\ trough is 17, with a range between
8 and 24.)

Considering Figure \ref{f_Cerrtau0300B} in more detail, we see that
in many individual velocity bins (each consisting of five original pixels)
both the minimum $C_v$ solution (thin red line) and the $C_v=1$ solution 
are formally acceptable at the 68\% confidence level ($\pm1\sigma$).
However, the lower panel of Figure \ref{f_Cerrtau0300B} 
shows that overall the
covering factor in the unblended outflow is statistically inconsistent with
both limiting cases.  Neither $C_v=1$ at all velocities nor $C_v$ having
its minimum value at all velocities are good fits to the data as a whole.
The upper panel corroborates the conclusion that the uniform $C_v=1$ solution
is unacceptable:  that solution corresponds to $\tau_v$ having its
minimum value at all velocities, whereas the best-fit optical depth is
consistently larger than the minimum.

\subsubsection{Broad \caii\ Absorption: Velocity Regime A} \label{VRA}

To model the \CaK\ absorption in velocity regime A (4337-5657~\kms),
we extrapolate from the absorption at 4000-4337~\kms\ in velocity regime B.
The absorption there, while noisy, is much more consistent with the minimum
$C_v$ than with $C_v=1$, and the optical depth is consistent with being
constant at its weighted average of $\tau_v=0.88\pm0.03$.  
We adopt a fixed $\tau_v=0.88$ for regime A, which means that the
corresponding fractional absorption in \CaH\ 
is 61\% of that in \CaK\ 
and that the covering factor $C_v$ is 1.71 times its minimum possible value
at each $v$.
We also calculate the limiting minimum-$C_v$ and $C_v=1$ cases, as they
delimit the range of plausible systematic errors.

\subsubsection{Broad \caii\ Absorption: Velocity Regime C} \label{VRC}

\begin{figure} \epsscale{1.0} 
\plotone{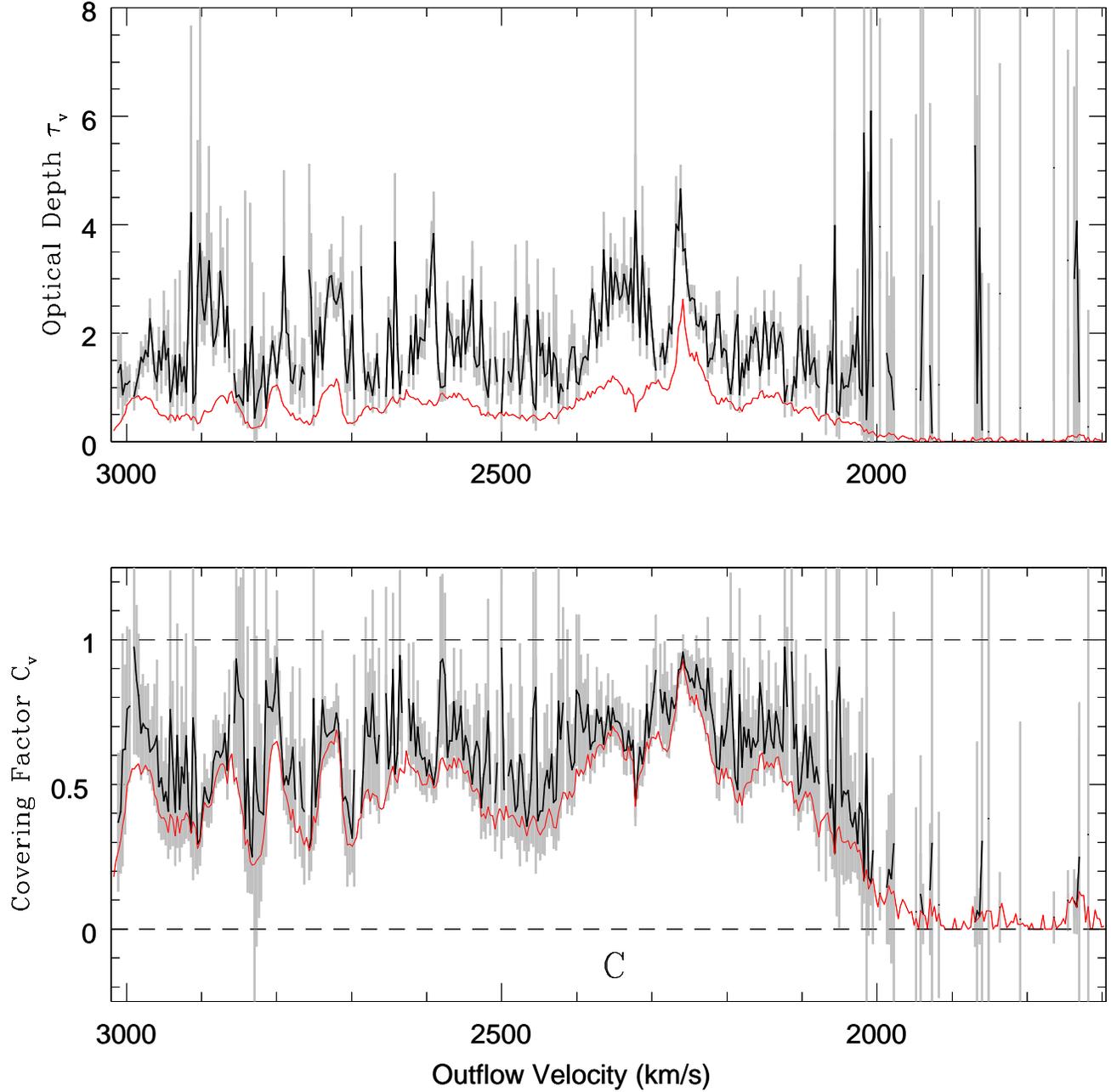}		
\caption[]{ \singlespace 
Optical depth (top)
and covering factor for \ion{Ca}{2} as a function of velocity
in the (blended) velocity regime C,
after binning by 5 pixels in the dispersion direction. 
Symbols as in Figure \ref{f_Cerrtau0300B}.
The uncertainty ranges (grey areas) include the systematic uncertainties 
from the assumptions made in order to remove the blended absorption, but
the random errors almost always dwarf the systematic uncertainties.
See \S\,\ref{VRC} for discussion.
}\label{f_Cerrtau0300C}
\end{figure}

To the observed residual \CaK\ intensities in this velocity range,
we add back in the fractional absorption in high-velocity \CaH\ estimated
in the previous section.  We then solve for $C_v$ and $\tau_v$ as per usual.
Figure \ref{f_Cerrtau0300C} shows the results.
The uncertainty ranges shown as the grey areas include the systematic
uncertainties arising from the removal of the high-velocity \CaH.
Note 
that the removal of high-velocity \CaH\ resulted in a large
increase in the number of pixels in this velocity regime 
with valid physical solutions.

In most velocity bins, the minimum $C_v$ solution cannot be ruled out but the
$C_v=1$ solution can.  On average, however, $C_v$ is significantly larger than
the required minimum value.  This behavior is similar to that in velocity
regime B, though not as pronounced.

\subsubsection{Broad \caii\ Absorption: Summary} \label{CAII}

\begin{figure} \epsscale{1.0} 
\plotone{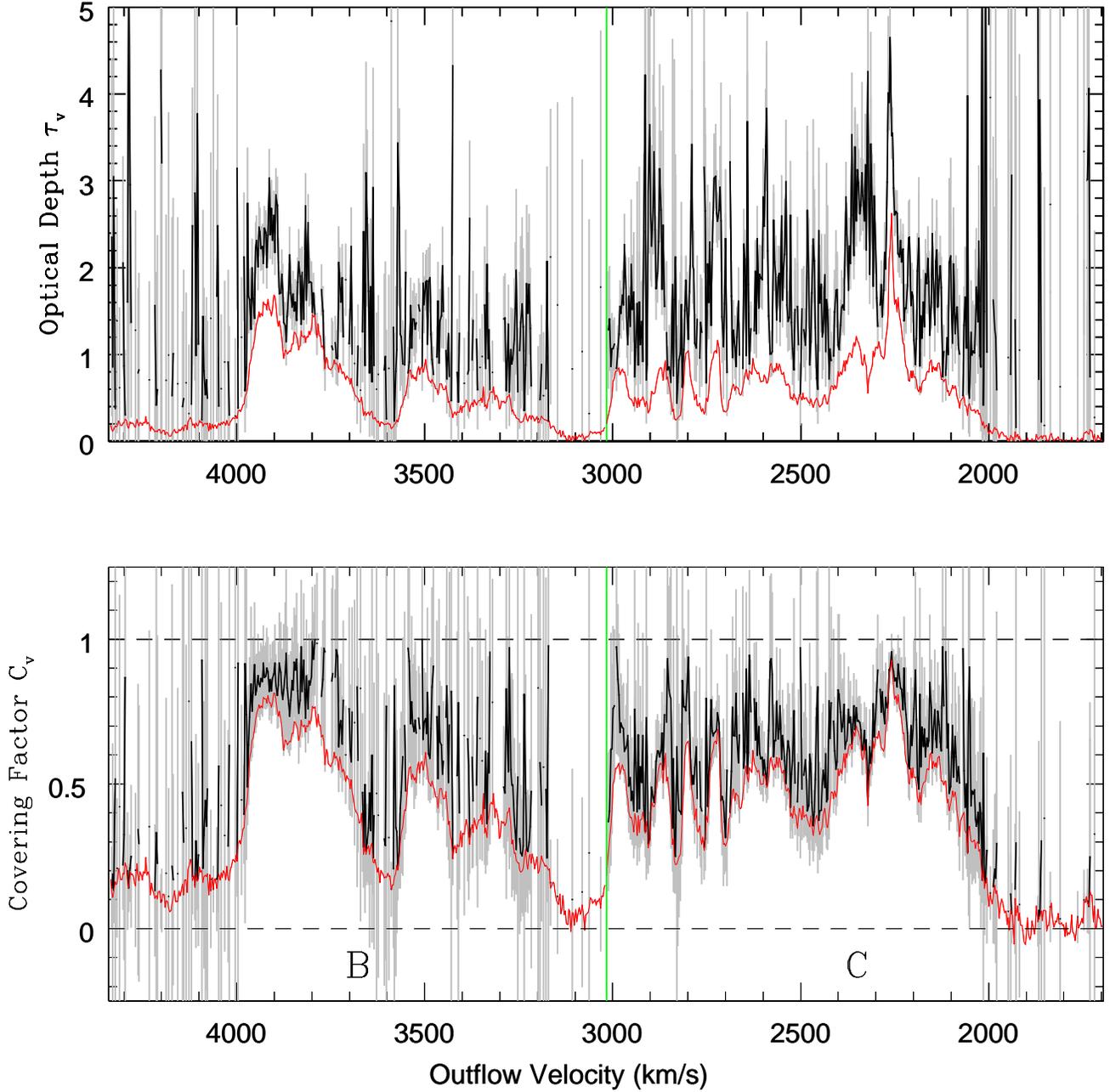}		
\caption[]{ \singlespace 
Optical depth (top) and covering factor as a function of velocity
for the \ion{Ca}{2} absorption,
after binning by 5 pixels in the dispersion direction.
Note that the optical depth scale is different than in the previous two Figures.
The green vertical dot-dashed line at 3017~\kms\ separates velocity
regimes B and C; other symbols as in Figure \ref{f_Cerrtau0300B}.
See \S\,\ref{CAII} for discussion.
}\label{f_Cerrtau0300BC}
\end{figure}

Figure \ref{f_Cerrtau0300BC} 
shows the optical depth and covering factor over the 
velocity range of strong \caii\ absorption.
As mentioned previously, the covering factor is 
on average significantly greater than the minimum value, though it tracks
the minimum value more closely in velocity regime C than in regime B.
The \caii\ outflow is moderately saturated ($1<\tau<3$), 
with only one significant spike above 4.
Thus, in general, the covering factor determines much --- but not all --- of 
the shape of the absorption profile in the \caii\ outflow.

\subsubsection{The Broad \MgI\ Absorption} \label{MGI}

\begin{figure} \epsscale{1.0} 
\plotone{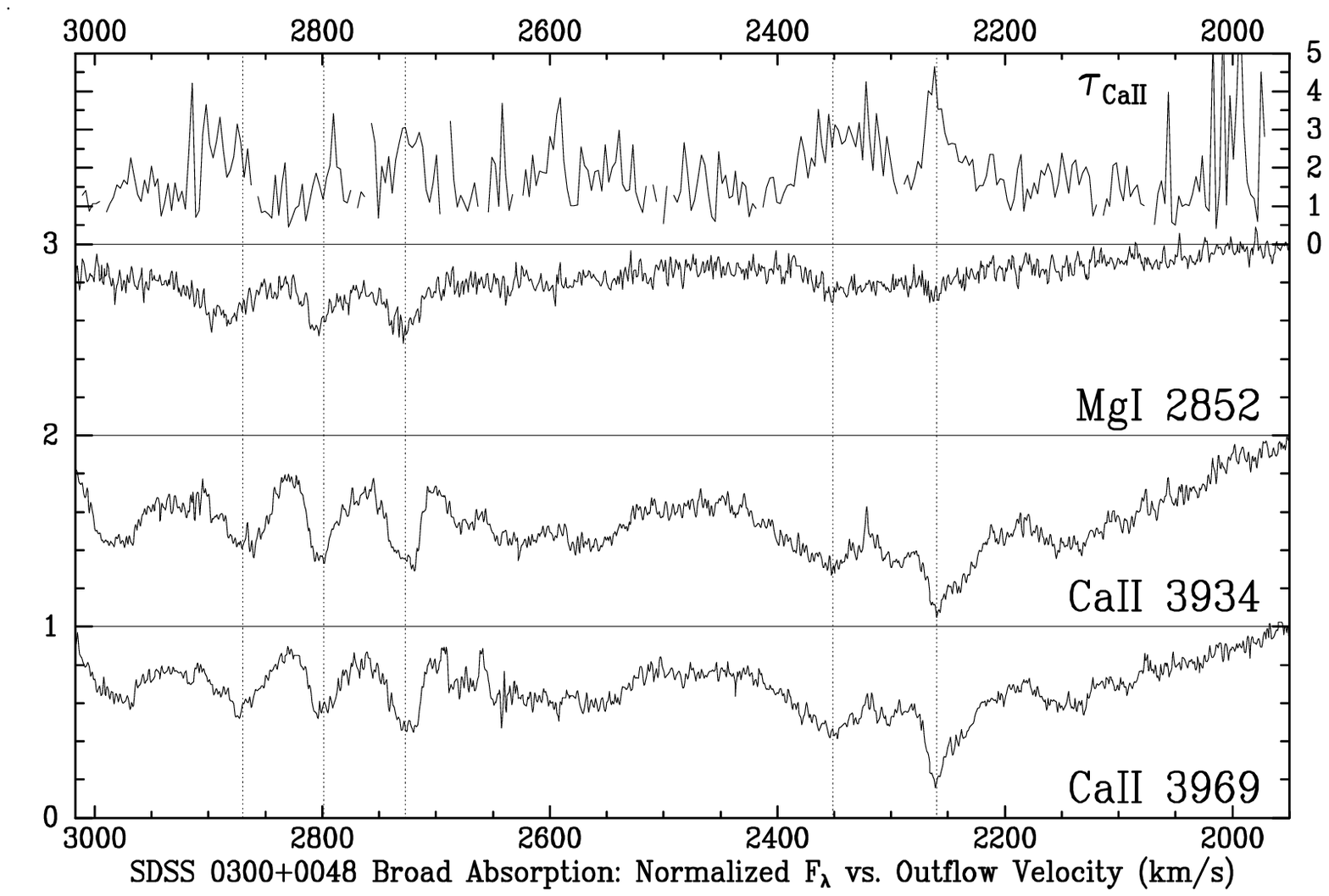}
\caption[]{ \singlespace 
Close-up of velocities where strong \mgi\ absorption is present in the BAL
outflow.  The uppermost panel and right-hand scale show the \CaK\ optical depth.
The lower three panels show normalized intensities in the \MgI, \CaK,
and \CaH\ regions, the latter corrected for narrow, blended associated \CaK.
The \mgi\ region normalization is uncertain,
as it is unclear whether there is \mgi\ absorption at 2400-2700~\kms.
See \S\,\ref{MGI} for discussion.
}\label{f_broad5wayzoom}
\end{figure}

As seen in Figure \ref{f_broad5way}, the strongest \mgi\ absorption in the broad
outflow is present between 2000-3000~\kms.
Figure \ref{f_broad5wayzoom} shows the absorption and \caii\ optical depth
over this velocity range in detail.  
Not all the fluctuations in $\tau_v$ are real, of course, but the five 
high-$\tau$ features marked with the dotted lines are statistically significant.
(The lone high-$\tau$ feature near 2600~\kms\ is likely a residual from the
correction of blended associated absorption in the \CaH\ trough.)
The two features between 2200-2400~\kms\ have weak \mgi.
The three narrow absorption features between 2700-2900~\kms\ have a larger
\mgi/\caii\ ratio because they have equal or smaller $\tau_{\rm CaII}$ (and
similar $\sigma_{\tau}$) but stronger \mgi\ absorption.\footnote{The
three features are seen in both \mgi\ and \caii, but the velocity distributions
of their absorption strengths differ in two features. 
The $\tau_{\rm CaII}$ does match the depths of the \mgi\ features rather well,
however, suggesting that \mgi\ is unsaturated while the large $\tau_v$ and
variable $C_v$ in \caii\ shift its velocity centroid away from that of \mgi.}
 
Note that there is no evidence for any features in the (heavily saturated)
\mgii\ trough at these velocities, nor is there any sign of
\cai~$\lambda$4227.9 absorption.  As with those in the associated absorber
(Appendix \ref{APPABS}), these \mgi/\caii\ ratio variations seem likely
to be due to metallicity or abundance variations, though
CLOUDY modeling is needed to firmly rule out an ionization parameter effect.
The striking similarity of the features in the 
\mgi\ and \caii\ troughs rules out the possibility that the two ions are
found in physically distinct regions that happen to share the same velocity.

A similar effect may be present in QSO J2359$-$1241. Table 1 of \markcite{aea01}{Arav} {et~al.} (2001) shows
that its absorption component $e$ has a \caii/\mgi\ ratio several times larger
than that in components $b$ and $c$.  Unlike in SDSS J0300+0048, the broader
component ($e$) has the higher \caii/\mgi\ ratio.  
However, the spectra of \markcite{aea01}{Arav} {et~al.} (2001) 
included only one line of the \caii\ doublet, so the reality of the
\caii/\mgi\ variation in QSO J2359$-$1241 remains to be confirmed.

\subsubsection{Column Densities in the BAL Outflow} \label{COLDEN}

For an ion giving rise to unsaturated and instrumentally well-resolved
doublet absorption,
the optical depth at velocity $v$ is directly proportional to
the column density per unit velocity
at $v$, $N_v$, which has units of cm$^{-2}$~(\kms)$^{-1}$ \markcite{ss91}({Savage} \& {Sembach} 1991).
The total column density \markcite{aea99}({Arav} {et~al.} 1999) and its uncertainty, both in units of
cm$^{-2}$, are then given by
\begin{equation}                 \label{e_N}
N = \int N_v dv = \sum N_v \Delta v = {{3.7679 \times 10^{14} }\over{\lambda_1 f_1}} \sum \tau_v \Delta v;
~~~\sigma_N = {{3.7679 \times 10^{14} }\over{\lambda_1 f_1}} \sqrt{\sum (\sigma_{\tau_v} \Delta v)^2}
\end{equation}                 
where we have ignored the uncertainties on the wavelength $\lambda_1$ (in \AA)
and the dimensionless oscillator strength $f_1$ 
of the stronger line of the doublet, which has optical depth $\tau_v$
in the velocity bin of width $\Delta v$ (in \kms) 
at velocity $v$.\footnote{$N$ and $N_v$ are independent of the covering factor
$C_v$, but the total mass present in that ion along the line of sight 
does depend on $C_v$.} 
For saturated absorption, Eq. \ref{e_N} yields a lower limit to the true $N$.
For a singlet transition with $C_v=1$,
$\sigma_{\tau_v} = \sigma_1 e^{\tau_v} = - \sigma_1/I_1$.
In our calculations of $N$ for various ions,
at velocities where $\tau_v$ or $\sigma_{\tau_v}$ are formally undefined,
we linearly interpolate using the two nearest defined values.
For singlet transitions, for velocities where $\tau_v$ is formally undefined we
assume that $\tau_v=-\ln(I_1)$, even if the result is a negative $\tau_v$.  

We are interested in the column densities in three different velocity regimes,
which are not quite the same as the \caii\ regimes ABC.
First is 1697-3740~\kms, where the \caii\ absorption is strong but excited
\feii\ absorption is absent.
Next is 3740-5657~\kms, where both \caii\ and excited \feii\ absorption are
present.
Last is 5657-8221~\kms, where \caii\ is absent but \mgii\ and excited \feii\ 
remain relatively unconfused.
The column densities we derive for the various ions in these velocity regimes
are listed in Table~\ref{t_colden} and are discussed below.
We do not measure column density limits for the additional $\sim$2600\,\kms\ of 
\mgii\ and excited \feii\ absorption present at $>$8221\,\kms\ 
(\S\,\ref{OVER0300}).

We find a total $N_{\rm CaII}=(7.13\pm1.15) \times 10^{14}$ cm$^{-2}$.
For \cai\,$\lambda$4227.9 we assume $C_v=1$ to set a limit of
$N_{\rm CaI} \leq 1.7 \times 10^{12}$ cm$^{-2}$ for the entire outflow.

We set a rough lower limit on $N_{\rm MgII}$ by ignoring 
\mgii\,$\lambda$2803 and summing only \mgii\,$\lambda$2796,
assuming $C_v=1$ and thus $\tau_v=-\ln(I_v)$, where $I_v$ is the residual
intensity at velocity $v$.  This yields 
$N_{\rm MgII} \geq (6.61 \pm 0.36) \times 10^{15}$ cm$^{-2}$
(for 1697-8221~\kms).
Normalizing by the strong \feii\ emitter continuum shown in Figure
\ref{f_0300fe2fit}, which is $\sim$2 times larger than the continuum we have
assumed, would increase $\tau$ by only $\sim\log(2)$ and 
$N_{\rm MgII}$ by only $\sim$20-25\%.  
However, because the \mgii\ trough is strongly saturated,
the value of $N_{\rm MgII}$ is a lower limit regardless of the continuum chosen.

\begin{figure} \epsscale{1.0} 
\plotone{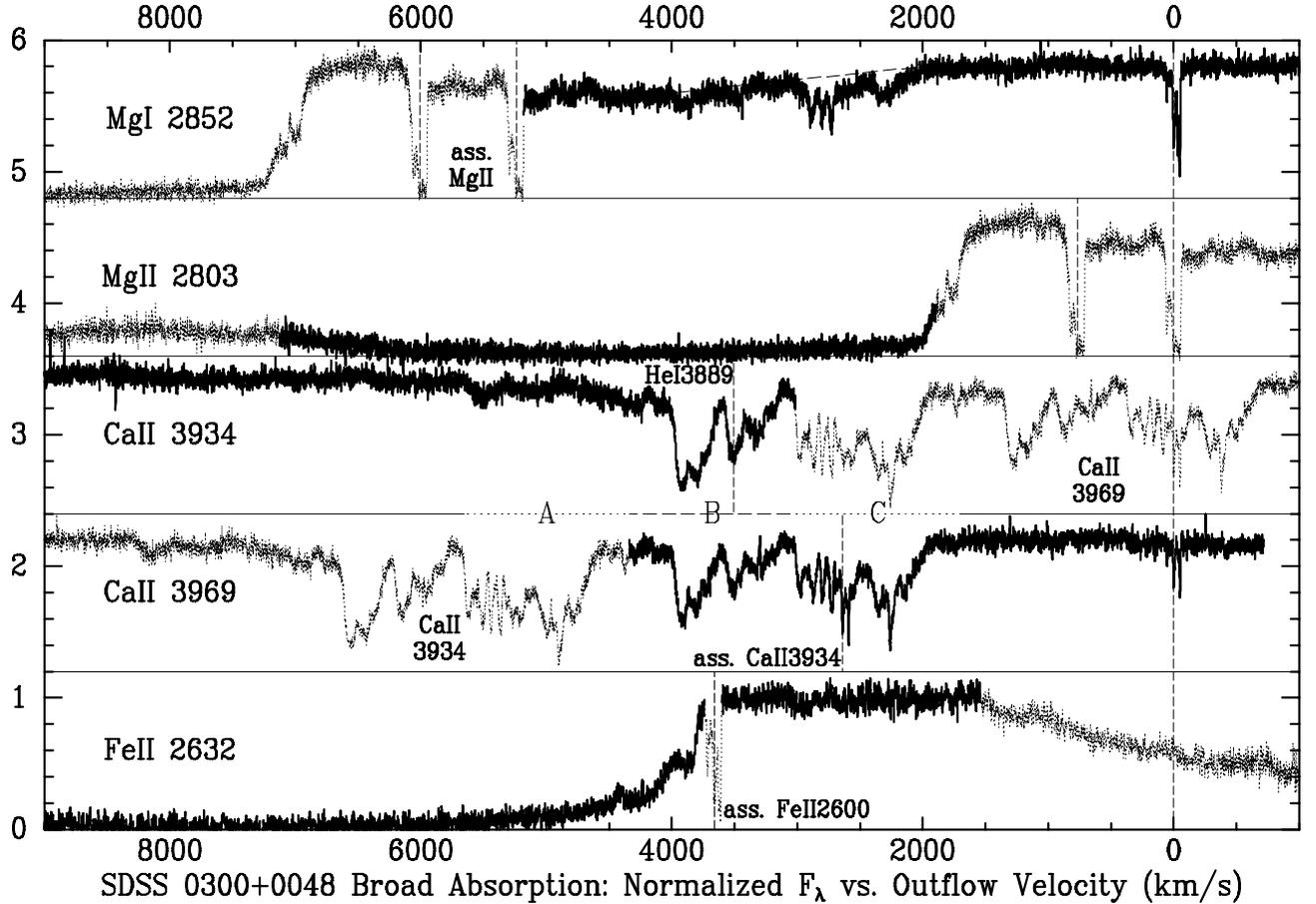}
\caption[]{ \singlespace 
The normalized spectrum of SDSS J0300+0048 around the broad absorption troughs
of five separate transitions. 
The spectrum of each transition is offset 1.2 units vertically from the one
below it, with the thin solid lines showing the zeo levels for each transition.
Unconfused regions of the spectrum in each transition
are shown with heavy lines. 
The velocity zeropoint is taken to be the redshift of the central component
of the associated absorption system, and outflowing velocities are positive; 
thus, shorter wavelengths appear on the left. 
Velocity regimes A, B and C are indicated between
the \CaK\ and \CaH\ panels (see \S\,\ref{BAL0300}).
Dashed vertical lines show the
locations of associated absorption; note that \feii\,$\lambda$2600 but not
\feii\,$\lambda$2632 is seen in the associated absorber.  The tilted dashed
line in the \mgi\ panel shows an alternate continuum normalization.
}\label{f_broad5way}
\end{figure}

For \mgi\ we set a conservative lower limit 
by assuming $C_v=1$, normalizing by the dashed line in
Figure \ref{f_broad5way}, and summing over $v$=2000-4500~\kms.
This yields $N_{\rm MgI} \geq (1.73\pm0.02) \times 10^{13}$ cm$^{-2}$.
A more liberal estimate, obtained using the default normalization shown in 
Figure \ref{f_broad5way}, 
yields $N_{\rm MgI} \geq (5.66\pm0.03) \times 10^{13}$ cm$^{-2}$
for 1697-5157~\kms, beyond which \mgi\ is confused with \mgii.  Alternatively,
using the strong \feii\ emitter continuum would increase the lower
limit on $N_{\rm MgI}$ by $\Delta N_{\rm MgI} = 1.1 \times 10^{14}$ cm$^{-2}$.
Clearly, the systematic uncertainty in the continuum placement is the
major uncertainty.

As for excited \hei\ (\hei* for short),
there is a possible detection of weak ($\lesssim 5$\%
of the continuum), broad \HeI\ absorption, but this line lies atop a broad
emission complex and so the continuum normalization is quite uncertain.
Also, despite confusion with \CaK, we see no sign of \HEI\ absorption with
depth more than $\sim5$\% of the continuum (Fig. \ref{f_broad5way}), 
even though its optical depth is three times that of \HeI.
\hei\,$\lambda$2945 provides no useful constraints because the observed
continuum surrounding it is simply too complex to be normalized accurately.
It is possible there is saturated \hei* absorption from a region with
$C_v \lesssim 0.05$, but we are most interested in limits on \hei* at velocities
where \caii\ or \mgii\ absorption is seen, and such absorption has 
$C_v \gg 0.05$.  To estimate a limit on $N_{\rm HeI^*}$, we assume absorption
5000~\kms\ wide with $C_v=1$, $\tau_{3188}=0.02$ and $\tau_{3889}=0.06$,
which yields $N_{\rm HeI^*} \leq 4.6 \times 10^{14}$ cm$^{-2}$.  

Our upper limit on the excited \feii\ column (\feii* for short)
at 1697-3740~\kms\ comes from
the non-detection of \ion{Fe}{2}\,$\lambda$2632, which is a blend of two lines.
Using NIST database $g_i f_i$ values for these lines from \markcite{dek01}{de Kool} {et~al.} (2001),
we find $N_{\rm FeII^*} \leq (0.42 \pm 1.62) \times 10^{13}$ cm$^{-2}$.
Our lower limits on the excited \feii\ column at higher velocities
come from assuming $C_v=1$ and attributing the absorption to
a blend of all lines from \ion{Fe}{2} multiplet UV1 at 2618-2633\,\AA,
a velocity span of 1600\,\kms.
We used NIST database $g_i f_i$ values from \markcite{dek01}{de Kool} {et~al.} (2001) for those lines also.
An exact accounting of which lines contribute at which wavelengths would
change the calculated lower limit somewhat, but the saturated absorption
means the column density is probably much greater than the lower limit anyway.
We do not attempt to convert these limits on the \ion{Fe}{2} column in
states $\sim$0.1\,eV above ground to limits on the total \ion{Fe}{2} column.

The above column densities could be underestimates if features in the outflow 
are barely resolved at our velocity resolution.  If this is the case,
recalculating the column densities after binning the spectrum will yield
significantly smaller values.  We find this to be only a $\sim$10\% effect:
the \caii\ column densities calculated from the spectrum after binning by five
pixels are about 15\% lower than the unbinned values in velocity regime B
and 5\% lower in velocity regime C.

\section{DISCUSSION}  \label{DISC}

\subsection{Broad \caii\ Absorption vs. Broad, Excited \feii\ Absorption}  \label{CAFE}

It is worth reiterating the interesting fact that the BAL region responsible
for the main \caii\ trough also produces \mgi\ and \mgii\ absorption, but none
from excited \feii\ except for a small velocity range
at the highest velocities where \caii\ is seen 
(\S\,\ref{OVER0300} and Fig. \ref{f_broad5way}).  Presumably the \caii\ region
does produce ground-state \feii\ absorption, but such absorption cannot be
studied because it is blended with higher-velocity excited \feii.

\begin{figure}
\epsscale{1.1}
\plotone{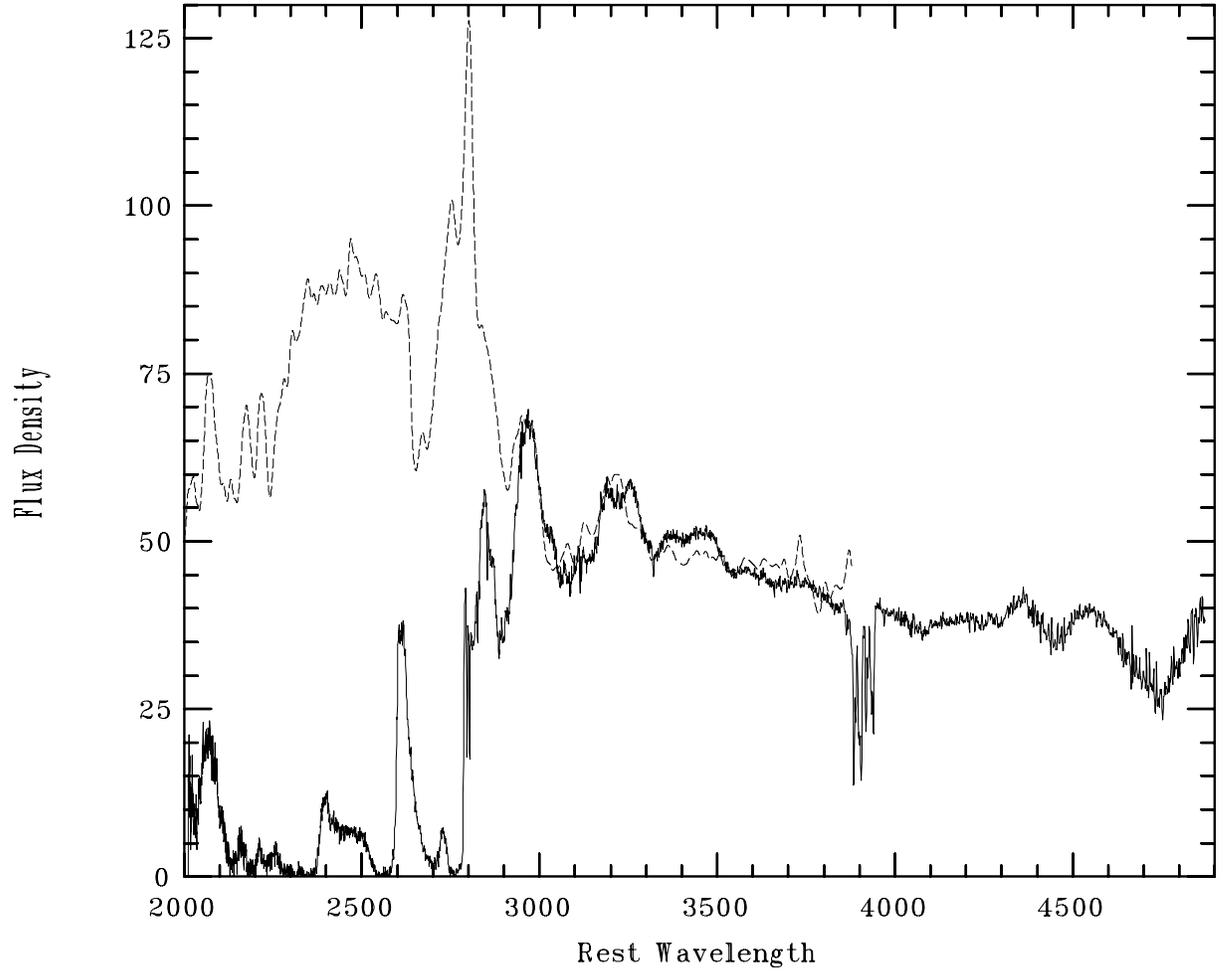}
\caption[]{ \singlespace 
Flux density $F_{\lambda}$, in units of
$10^{-17}$\,ergs cm$^{-2}$\,s$^{-1}$\,\AA$^{-1}$,
of SDSS J0300+0048 (solid line)
and a model for its intrinsic continuum (dashed line); see \S\,\ref{WHAT}).
}\label{f_0300fe2fit}
\end{figure}

Because the true continuum at the wavelengths where we expect to see excited
\feii\ absorption from the \caii-absorbing gas is likely much higher than we
have assumed (Fig. \ref{f_0300fe2fit}), there is almost certainly some
absorption present at those wavelengths.  Nonetheless, we can rule out the
possibility that excited \feii\ is present at 2000-3500~\kms\ with nonblack
saturation.  For that to be the case,
the unabsorbed continuum at the velocities in question would have to be higher
than we have assumed by a factor equal to the inverse of the partial covering.
The \caii\ absorber reaches $C_v \simeq 0.95$ in this velocity range,
thus requiring for this scenario an unabsorbed continuum twenty times higher
than assumed.  However, only a factor of two higher continuum
is expected at rest 2600\,\AA\ even if the
true continuum is that of a strong \feii\ emitter (Fig. \ref{f_0300fe2fit}).

One possible explanation 
for the different velocity distributions of \caii\ and excited \feii\ is that
the \caii\ region has a density too low for collisional excitation to
significantly populate excited states 
of \feii\ 
($n_e < 10^3$~cm$^{-3}$; \markcite{dek01}{de Kool} {et~al.} 2001);
at such densities radiative excitation will not contribute significantly either
(\S\,4.1.3 of \markcite{wcp95}{Wampler}, {Chugai}, \& {Petitjean} 1995).
Alternatively, the \caii\ could arise from a region with a density similar to
that in the \feii\ region, but with a lower temperature.
The levels that give rise to \feii\,$\lambda\lambda$2632 absorption
have an average excitation potential 
of 0.095\,eV, so a temperature $T \lesssim 1100$~K is required 
to avoid populating them significantly.  
Absorption from the somewhat weaker \feii\,$\lambda$2626 line is not detected
either, and its excitation potential is only 0.048\,eV.  Thus the temperature
could be as low as $T\lesssim550$~K in the \caii\ region.
Which of these alternatives 
--- low density or low temperature ---
is most likely the case for the \caii\ region
depends on the structure of the outflow, as we now discuss.

\subsection{Column Densities and Outflow Models} \label{CODIS}

Compared to QSO J2359$-$1241 \markcite{aea01}({Arav} {et~al.} 2001), the only
other \caii\ BAL quasar studied at high resolution, SDSS J0300+0048 has lower
limits on the total \mgi\ and \mgii\ column densities $\sim$30 times larger,
a lower limit on the excited \feii\ column $\sim$95 times larger,
an upper limit on the excited \hei\ column $\sim$8 times 
larger, and a total \caii\ column $\sim$200 times larger.
The column densities for all but the excited \feii\ are the largest reported
to date in any BAL outflow.

These values are consistent with the total column density of metals being 
$\sim$200 times larger in SDSS J0300+0048 than in QSO J2359$-$1241.
The excited \hei\ column need not increase as much because it comes from the
\heii\ region instead of the low-ionization BAL region that produces most of
the absorption in SDSS J0300+0048.  Also, the excited \hei\ absorption can be
greatly reduced if the \heii\ region has an $n_e$ 
below the critical density for this excited state ($3\times10^3$~cm$^{-3}$).

\markcite{aea01}{Arav} {et~al.} (2001) found that the absorption in 
QSO J2359$-$1241 could be explained
by solar metallicity gas with $\log N_{\rm H} \simeq 20$ and $\log U\simeq -3$.
The existence of a hydrogen ionization front requires a column density
$N_{\rm H} \simeq U \times 10^{23}$ cm$^{-2}$ 
\markcite{dn79}({Davidson} \& {Netzer} 1979), or a few times larger than that
for electron densities $n_e \gtrsim 10^{9.5}$ cm$^{-3}$
\markcite{dek02b}({de Kool} {et~al.} 2002a).  QSO J2359$-$1241 has just enough
column density to have a hydrogen ionization front ($\tau \sim 2.5$ at the
Lyman limit).  The $\sim$200 times larger \caii\ column density in 
SDSS J0300+0048 means that it must have a strong hydrogen ionization front.
Even if SDSS J0300+0048 is one of the most metal-rich quasars known, with
metallicity $\sim15$ times solar \markcite{bea03}({Baldwin} {et~al.} 2003),
it would have to have a \caii\ region with $N_H \sim 13$ times larger than in
QSO J2359$-$1241, corresponding to $\tau\sim 33$ at the Lyman limit.

The presence of an \hi\ ionization front in SDSS J0300+0048 agrees with
expectations from photoionization calculations 
\markcite{fp89}({Ferland} \& {Persson} 1989).
An \ion{H}{1} ionization front surrounding the quasar\footnote{The
ionization front is centered on the quasar, so we use the terms
{\em inside} and {\em outside} the front instead of {\em in front of} 
and {\em behind} the front.} can shield \mgii\ and \feii\ from ionization
\markcite{vwk93}({Voit}, {Weymann}, \& {Korista} 1993).  However, such a front
by itself cannot protect \mgi\ or \caii:  because those ions have lower 
ionization potentials than \hi, they are only found where the ionization 
parameter is low \markcite{hea01}({Hamann} {et~al.} 2001).
\mgi\ can be found in low-ionization gas with total column density less
than, or just equal to, that needed for an \hi\ ionization front
\markcite{aea01}({Arav} {et~al.} 2001).
\caii\ cannot, due to resonant photoionization from the lowest metastable level 
of \caii\ by \lya\ photons \markcite{jol89}({Joly} 1989).
Thus, \caii\ only exists well outside an \hi\ front, where not just the
ionization parameter but also the densities of \lya\ photons and photons
with $E>11.9$\,eV --- required to ionize \caii\ --- are all low.
In fact, for the clouds of density $10^{9.5}$\,cm$^{-3}$ modeled by
\markcite{fp89}{Ferland} \& {Persson} (1989), \caii\ 
only becomes the dominant ionization stage of calcium 
behind the \ion{C}{1} ionization front, at a column density
more than a factor of ten higher than that of the \hi\ ionization front.

This requirement has implications for the structure of the outflow, because it
means that the excited \feii\ BAL region cannot be farther from the quasar
than the \caii\ BAL region.  If it was, it would lie outside the \hi\ 
ionization front and would also show strong \caii\ absorption.  
(The weak \caii\ absorption from the excited \feii\ region means that at most
$\lesssim$10\% of the excited \feii-absorbing gas could lie outside the front.)
This fact, along with the
observations that the broad absorption is detached from the systemic redshift
and that the \caii\ BAL region lies at low velocities (1697-3740~\kms) while the
excited \feii\ BAL region lies at high velocities (3740-8221~\kms\ and beyond),
enable us to constrain which of two representative, competing BAL outflow
models might explain SDSS J0300+0048.  These models are the disk wind model of
\markcite{mur95}{Murray} {et~al.} (1995) and a model where the absorption
arises in confined clouds embedded with very small filling factor in a less
dense, hotter medium \markcite{al94}(e.g., {Arav} \& {Li} 1994).
Such clouds could form a dusty outflow that nearly 
surrounds the quasar \markcite{bea00}(e.g,. {Becker} {et~al.} 2000),
or they could be `ablated' from denser sources, such as an accretion disk 
or obscuring torus, and only block some lines of sight to the quasar.


In the radiatively driven disk wind model of 
\markcite{mur95}{Murray} {et~al.} (1995), the streamlines of the outflow begin
at some angle to the disk and then bend to become asymptotically radial (in
radial cross-section).  Thus the detachment velocity simply reflects the height
above the disk reached by the flow before it crosses the line of sight
to the continuum source; that height could be different for different ions.
As the distance from the quasar increases {\em along a single streamline},
the velocity and ionization increases and the density decreases.
For {\em two} streamlines arising at {\em different} initial radii,
the more distant streamline will be scaled up in density 
(\markcite{ss73}{Shakura} \& {Sunyaev} 1973)
but down in velocity and ionization parameter
\markcite{mur95}{Murray} {et~al.} (1995).
Because we see a detached flow in SDSS J0300+0048,
in this model our line of sight must cross numerous streamlines.
Thus, we expect to see higher-velocity gas that is
less dense, more highly ionized, and located at smaller radii.\footnote{The
simulations of \markcite{psk00}{Proga}, {Stone}, \& {Kallman} (2000) verify
these analytic predictions of \markcite{mur95}{Murray} {et~al.} (1995)
for radiatively driven disk winds.  These predictions are less secure in the
magnetohydrodynamic simulations of \markcite{pro03}{Proga} (2003), which yield
a turbulent zone of considerable solid angle
between the fast radiatively-driven wind component
and a slower MHD component that develops closer to the equatorial plane.
Should the line of sight pierce that latter zone, the relationship between
velocity, density, ionization parameter and distance may break down.}

The BAL outflow in SDSS J0300+0048 is consistent with such a disk wind, as the
lowest-ionization gas (traced by \caii\ and \mgi) is found at the lowest
velocities.
To be consistent with the observed anticoincidence of excited \feii\ absorption
and \caii\ absorption requires that a disk wind in SDSS J0300+0048 have
$n_e > 10^3$ cm$^{-3}$ throughout but $T\lesssim1100$~K at the low velocities 
(large radii) where \caii\ is present.
If the velocity decreases monotonically with radius along the line of sight,
then the \hi\ ionization front will shield gas at all velocities less than
some velocity $v_s$.  
Because the \caii\ absorption can only arise outside the \hi\ ionization front,
we surmise either that 
$v_s \simeq 5750$~\kms, which is the maximum observed \caii\ velocity, or that
$v_s \simeq 4000$~\kms, which divides strong and weak \caii\ absorption, and
that the higher-velocity \caii\ absorption is due to 
a small portion (by mass) of the outflow departing from 
a monotonically decreasing velocity. 
If the \mgii\ trough were not saturated, it might reveal the velocity of the
front, as the ionic fraction of \mgii\ increases sharply outside the front
\markcite{vwk93}({Voit} {et~al.} 1993).


If we try to explain the \caii\ absorption as arising from clouds 
with a lower density than the clouds 
giving rise to excited \feii\ absorption, but at similar radii,
then the \caii\ BAL clouds must have $n_e < 10^3$~cm$^{-3}$
while the excited-\feii\ BAL clouds have a higher $n_e$ and lower $T$,
assuming all clouds are in pressure balance with the same confining medium.
However, the column density in the \caii\ clouds must be larger than that in
the excited-\feii\ BAL clouds, which leads to an improbable scenario where 
diffuse clouds reach high column densities but dense clouds do not.
%
Thus, as in the disk wind case, we are driven to a scenario where
the excited \feii\ BAL clouds must have $n_e > 10^3$~cm$^{-3}$
while the \caii\ BAL clouds have an even higher $n_e$ but a temperature 
$T\lesssim1100$~K, to avoid populating the excited \feii\ levels.

For a cloud scenario where the outflow surrounds the quasar, 
the outflow must have smaller velocities at larger radii 
because the \caii\ BAL region must lie at larger radii 
than the excited \feii\ BAL region.
This velocity distribution might occur if this quasar recently `turned on'
inside a relatively thick shell or torus of gas, and has gradually radiatively
accelerated (and radially compressed)
it such that the velocity in the shell is a decreasing function of radius.
The \caii\ absorption would arise in the outermost regions of the flow, which
have only been accelerated to velocities of $\sim$2000-4000~\kms\ so far.
This scenario is rather contrived, especially as such an outflow will be
Rayleigh-Taylor unstable (though magnetic fields might suppress the
instabilities), but it can be tested by looking for a long-term increase
in the velocity of the \caii\ BAL outflow.

A cloud scenario where the outflow does {\em not} surround the quasar
can explain the detached troughs by positing that the injection points of the
clouds (where they have $v_{radial}=0$) do not lie along our line of sight
to the continuum source, and that there is some finite outer radius for the
injection that corresponds to the minimum radial velocity we observe.
Thus the outflow cannot be spherically symmetric,
as such a flow would have injection points along all lines of sight.
Injection of clouds from a disk or torus 
is still feasible.
The clouds must cross our line of sight in a spatially- and velocity-segregated
flow, as in the disk wind case, because the (low-velocity) \caii-absorbing 
clouds must be at larger radii than the (high-velocity) \feii-absorbing clouds.
If clouds of all velocities were intermixed, the gas in any high-velocity clouds
shadowed by a low-velocity \caii-absorbing cloud would recombine to lower
ionization states on timescales of $\sim10^{11}/n_e$ seconds.  The result would
be \caii\ absorption at high velocities as well as low, which is not seen.

Thus, for a cloud model to explain this outflow the clouds must form either a
rather contrived expanding shell or a segregated flow 
arising from a disk or torus.  The latter case differs from the disk wind model
only in the presence of an intermixed, hot, confining medium for the 
gas responsible for the low-ionization absorption.

Finally, it is not clear how the associated absorption fits in to either the
cloud or the disk wind model.  However, in both models it must arise far enough
from the continuum source that it has not been significantly accelerated
along our line of sight.  
Another of the five overlapping-trough BAL quasars in H02, SDSS J0819+4209,
has BAL troughs detached from an AAL system. 
That object has a similar detachment velocity ($v=2180\pm80$~\kms),
but the \mgii\ AAL appears to be broader than in SDSS J0300+0048.
Anecdotally, AALs and detached BALs appear to be common in additional
overlapping-trough quasars being found by the SDSS, but the possible
significance of this finding to models of these outflows remains unclear.

\section{Conclusion}  \label{CONCLUDE}

Comparison with QSO J2359$-$1241, the only other \caii\ BAL quasar studied at
high resolution, suggests that the outflow in SDSS J0300+0048 has a strong
\hi\ ionization front, in agreement with the photoionization modeling of
\caii\ by \markcite{fp89}{Ferland} \& {Persson} (1989).  The observed \caii\ 
absorption must arise in low-ionization gas located outside the ionization
front.  Kinematically, the outflow is detached --- it only begins at a velocity
blueshifted by $\sim$1700~\kms\ from the 
associated absorption redshift. 
Furthermore, excited \feii\ absorption is seen only at high velocities,
and strong \caii\ absorption only at the low velocity end of the outflow.

The velocity and ionization structure in the SDSS J0300+0048 outflow can be
explained in a disk wind model where the \caii\ arises only at large radii.
A cloud model for the BAL outflow can explain the observations if the clouds
form a segregated flow originating from a source out of our line of sight ---
in other words, a clumpy disk wind instead of a smooth one
\markcite{eve03}(e.g., {Everett} 2003).
An outflow of clouds that completely surrounds the quasar can explain the
observations only in the case of a shell of gas in which the velocity decreases
with increasing radius and in which the gas has been uniformly accelerated
such that none of it has velocities $v \lesssim 1700$~\kms.
While this model is rather contrived, it is testable because it predicts
a long-term increase in the observed velocity of the outflow.

The region producing excited \feii\ absorption must have $n_e > 10^3$~cm$^{-3}$
and $T\gtrsim 1100$~K to significantly populate the observed fine structure
levels.  In any of these models, it is almost certain that the region producing
\caii\ absorption also has $n_e > 10^3$~cm$^{-3}$, in which case the
\caii-absorbing region must have $T \lesssim 1100$~K --- and perhaps as low as
$T \lesssim 550$~K --- to prevent excited \feii\ states from being populated
there.

More generally, quasars with \caii\ troughs (and thus \hi\ ionization fronts)
which are detached in velocity enable a test to be made of the 
\markcite{mur95}{Murray} {et~al.} (1995) analytic disk wind model prediction
of a monotonically decreasing velocity with increasing radius
along the line of sight.
Because the lowest velocity absorption arises outside the front, no absorption
from high-ionization gas (e.g., \civ\ or \ion{N}{5}) should be present at the
lowest velocities.  (Undetached troughs are not useful for this test because
if the streamlines produce gas with zero velocity along the line of sight at 
one radius, they could easily produce such gas at many different radii.)
Unfortunately, this test cannot be made in SDSS J0300+0048 because \feii\ 
absorption obliterates the spectrum.  It might be possible with spectra covering
\civ\ in \caii\ LoBALs with narrower troughs, such as FIRST J104459.6+365605,
whose \caii\ absorption is found at the low-velocity end of its higher-velocity
outflow \markcite{wea00,dek01}({White} {et~al.} 2000; {de Kool} {et~al.} 2001).
On the other hand, the velocity distributions are the {\em same} for \caii,
excited \feii, and excited \hei\ absorption (which traces high-ionization \heii)
in the other two known \caii\ LoBALs:
QSO J2359$-$1241 \markcite{aea01}({Arav} {et~al.} 2001) and
Mrk~231 \markcite{smi95,rvs02}({Smith} {et~al.} 1995; {Rupke} {et~al.} 2002).
In fact, wide ranges of ionization states at identical velocities are often
seen in BAL outflows \markcite{aea99}(e.g., {Arav} {et~al.} 1999).
If disk wind models are to explain all BAL quasars, therefore, such models must
at least sometimes produce a nonmonotonic velocity-distance relation or density
inhomogeneities of some sort (such as clouds).

The associated absorption line system in SDSS J0300+0048 (see Appendix
\ref{APPABS}) exhibits partial covering, and therefore is likely located near
the central engine.  Whether it is associated with the BAL outflow is
unclear.  Blending of the AAL with the BAL trough shows that the spatial regions
covered by the BAL outflow can vary significantly over velocity differences of
$\sim$1700~\kms.  Variations in the \mgi/\caii\ ratio with velocity are also
seen in both the BAL and AAL systems; these appear more likely to be due to
variations in metallicity or abundances than in the ionization parameter.

In conclusion, in H02 we stated that ``targeted high resolution spectroscopy of
\caii\ in SDSS J0300+0048 ... might determine if the column densities as well
as the outflow velocities are very large."  We have shown herein that the BAL
outflow column densities are among the largest yet measured in BAL outflows.
Furthermore, the highly ionized gas columns are as yet largely unconstrained.
{\em Chandra} and possibly {\em XMM-Newton} observations will be able
to constrain the total BAL column density regardless of ionization state.

\acknowledgements

We thank F. Barrientos, C. Churchill, H. Dickel, I. Strateva, J. Veliz
and the referee.  P. B. H. acknowledges financial support from Chilean
grant FONDECYT/1010981 and from Fundaci\'{o}n Andes.
Funding for the creation and distribution of the SDSS Archive has been provided
by the Alfred P. Sloan Foundation, the Participating Institutions, the National
Aeronautics and Space Administration, the National Science Foundation, the U.S.
Department of Energy, the Japanese Monbukagakusho, and the Max Planck Society.
The SDSS Web site is http://www.sdss.org/.  The SDSS is managed by the
Astrophysical Research Consortium (ARC) for the Participating Institutions.
The Participating Institutions are The University of Chicago, Fermilab, the
Institute for Advanced Study, the Japan Participation Group, The Johns Hopkins
University, Los Alamos National Laboratory, the Max-Planck-Institute for
Astronomy (MPIA), the Max-Planck-Institute for Astrophysics (MPA),
New Mexico State University, University of Pittsburgh, Princeton University,
the United States Naval Observatory, and the University of Washington.
This research has made use of
the Atomic Line List v2.04 at \url{http://www.pa.uky.edu/$\sim$peter/atomic/}.

\begin{appendix}

\section{The Associated Absorption} \label{APPABS}

We are interested primarily in the broad \caii\ absorption, but there is
narrow associated \CaK\ blended with the broad \CaH.
To remove this absorption we must model the associated system.

To determine the continuum around each absorption line from the associated
absorber, we make the simplifying assumption
that the associated absorption is {\em external} to the broad absorption line
region, an assumption we reconsider at the end of this section.
Under this assumption, the locally measured continuum around each line is
a good estimate of the true continuum seen by the associated absorber.
Because there is structure near the associated \mgii\ that appears to be due to
additional weak \mgii\ emission or absorption, we construct a local continuum
for the associated \mgii\ by fitting a single polynomial continuum, offset
only by a small flux difference, around both transitions in velocity space.

Figure \ref{f_6way} shows that
the associated system is seen in \caii, \mgii, \mgi, \feii\,$\lambda$2600,
\feii\,$\lambda$2382 (and \feii\,$\lambda$2374, which is not plotted),
but not in \feii\,$\lambda$2586 (all these \feii\ lines are from the lowest
--- i.e., unexcited --- fine-structure level in the ground term of \feii).
We will return to this point at the end of this section. 
\CaH\ and \mgi\ absorption are both weaker than \feii\,$\lambda$2600
absorption, but their profiles are broadly consistent with the
\feii\,$\lambda$2600 absorption profile seen at lower optical depths.
Because both ions exist only in gas with a low ionization parameter,
they should show the same covering factors if they are saturated.
We conclude that the \caii\ and \mgi\ absorption lines are unsaturated,
with different $\tau_v$.
Overall, the associated absorption appears to have
a higher \mgi/\caii\ ratio than the broad absorption does (\S\,\ref{OVER0300}).
Also, within the associated absorption itself,
the subsystem at $-$32 \kms\ has a noticeably higher \mgi/\caii\ ratio.
This fact suggests the existence of inhomogeneities in the ionization,
metallicity, or relative abundances as a function of velocity in the
associated absorber (\markcite{cea03}{Churchill} {et~al.} 2003,
and references therein).

We can use the unsaturated associated \CaH\ absorption to 
model the optical depth in associated \CaK, but 
the \mgii\ absorption
indicates that the associated absorber 
exhibits partial covering of the continuum source.
Even though \mgii\ can occur in somewhat higher ionization gas than \caii\ can,
we can do no better than to assume that the associated \caii\ 
has the same partial covering with velocity as \mgii.

\begin{figure}
\epsscale{1.0}
\plotone{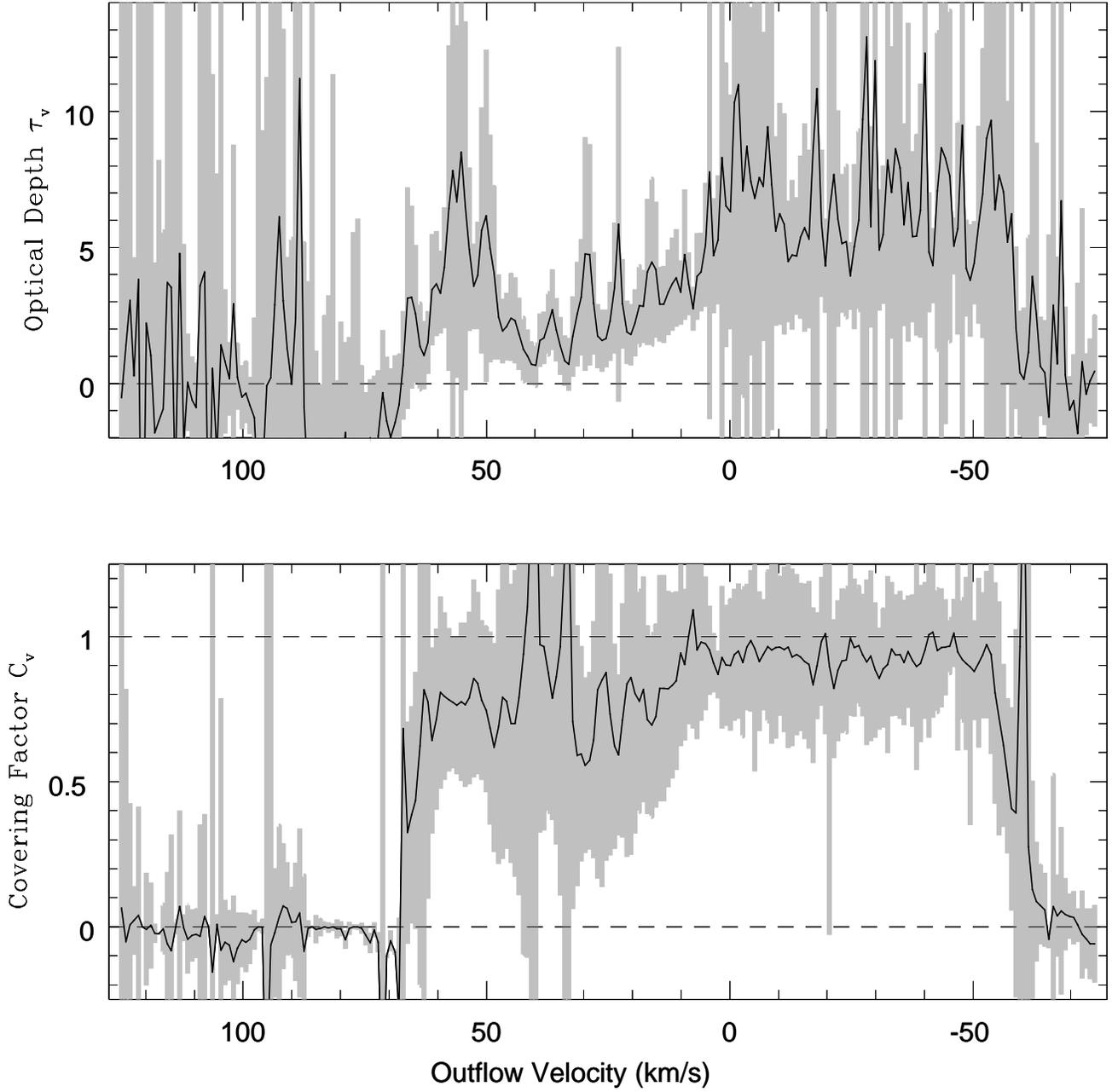}              
\caption[]{ \singlespace 
The thick black line gives the optical depth (top)
and covering factor for associated \mgii\ as a function of velocity.
The grey areas show the $\pm1\sigma$ uncertainty ranges.
The dashed lines delimit the ranges
of physically valid solutions; namely, $\tau_v > 0$ and $0\leq C_v \leq1$.
}\label{f_assfit}
\end{figure}

The results of using Eqs. \ref{e_C}-\ref{e_tau} to 
compute $C_v$ and $\tau_v$ for \mgii\ 
are shown in Figure \ref{f_assfit}.
The absorber has $\tau_v \geq 1$ at almost every velocity.
The observations are not quite consistent with $C_v=1$ at all velocities:
$C_v$ appears lower at $v>10$~\kms\ than at $v<10$~\kms.
Because the relative $C_v$ of \caii\ and \mgii\ are somewhat uncertain
and the observed variations in $C_v$ are consistent with the random noise,
we will approximate $C_v$ as a step function. 
From $-$54 to 10.5 \kms\ we find an average partial covering of $C=0.93\pm0.04$
(with $\tau_{\rm MgII} \simeq 7$). 
From 10.5 to 64 \kms\ we find an average partial covering of $C=0.76\pm0.09$
(with $\tau_{\rm MgII} \simeq 3$). 
There is additional absorption 
from 64 to about 93 \kms\ for
which the ($C,\tau$) fit breaks down because the 2796\,\AA\ absorption is less
than half the strength of the 2803\,\AA\ absorption.  It is likely that our
continuum fit is in error at these velocities in one or both of the \mgii\ 
lines. 

\begin{figure} \epsscale{0.66} 
\plotone{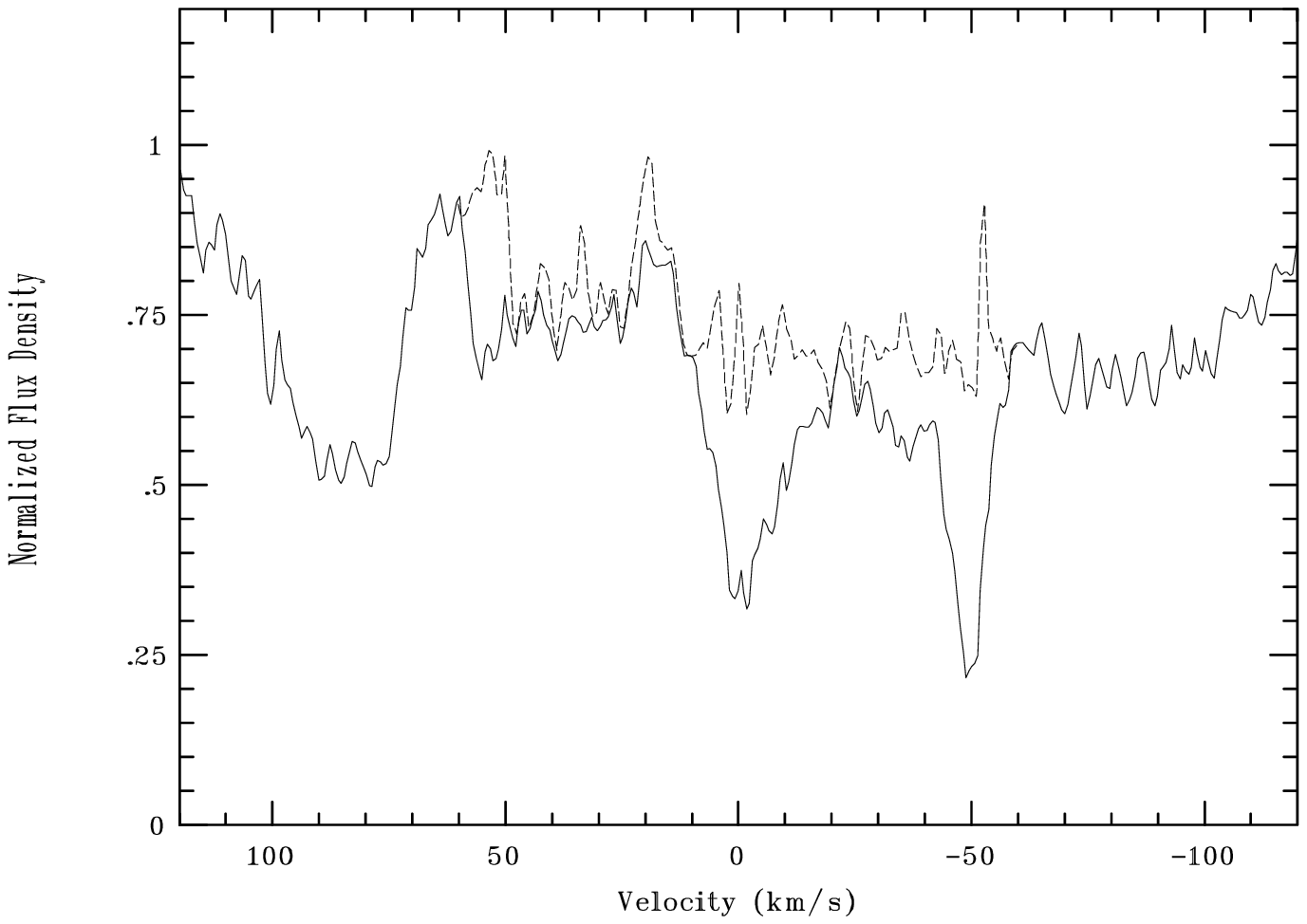}
\caption[]{ \singlespace 
The region of the \CaH\ BAL trough around the wavelength of associated 
\CaK\ absorption. 
The solid line shows the observed, normalized spectrum, and the dashed line
the spectrum with the associated \CaK\ removed using the \CaH\ profile and
estimates for the optical depth and covering factor; see Appendix \ref{APPABS}.
The corrected spectrum appears reasonable everywhere except
for a narrow spike near $-$53 \kms. 
}\label{f_OandF}
\end{figure}

We now apply these two covering factors to the associated \CaH\ absorption 
to predict the optical depth of the associated \CaK\ absorption.
If the associated absorption is external to the broad absorption, then
--- independent of any assumption
about the true continuum incident on the broad absorption line region ---
the flux corrected for the associated \CaK\ absorption $F$ 
is related to the observed flux $O$ by
\begin{equation}                 \label{e_obsflux}
O = F \times [1 - C \times (1 - e^{-R \tau}) ] 
\end{equation}
where $\tau$ is the optical depth in \CaH\ and all quantities except $R$ are
functions of velocity.  We can recover $F$ because we know $O$ and $R$ and
have estimates for $C$ and $\tau_{3969}$.
Figure \ref{f_OandF} shows the spectral region around associated \CaK\ before
(solid line) and after (dashed line) this correction.  The corrected spectrum
appears reasonable everywhere except for a narrow spike 
near $-$53 \kms, which is interpolated over for our analysis.

\subsection{Associated or Broad Absorption: Which Comes First?} \label{ASSD}

Because the associated \CaK\ spans such a narrow velocity range, our
correction for it is not critical to the subsequent analysis of the BAL trough.
Nonetheless, before proceeding it is worth considering further our assumption 
that the associated absorption arises outside the broad absorption line region.
The probable partial covering of the associated absorber immediately calls into
question this assumption, because partial covering is seen almost exclusively
in gas linked with the central engines of AGN. 
However, while we cannot rule out a close connection between the associated and
broad absorbers, we now show that neither can we rule out that the
associated absorption 
arises near the nucleus but external to 
the broad absorption region.

\begin{figure}
\epsscale{1.0}
\plotone{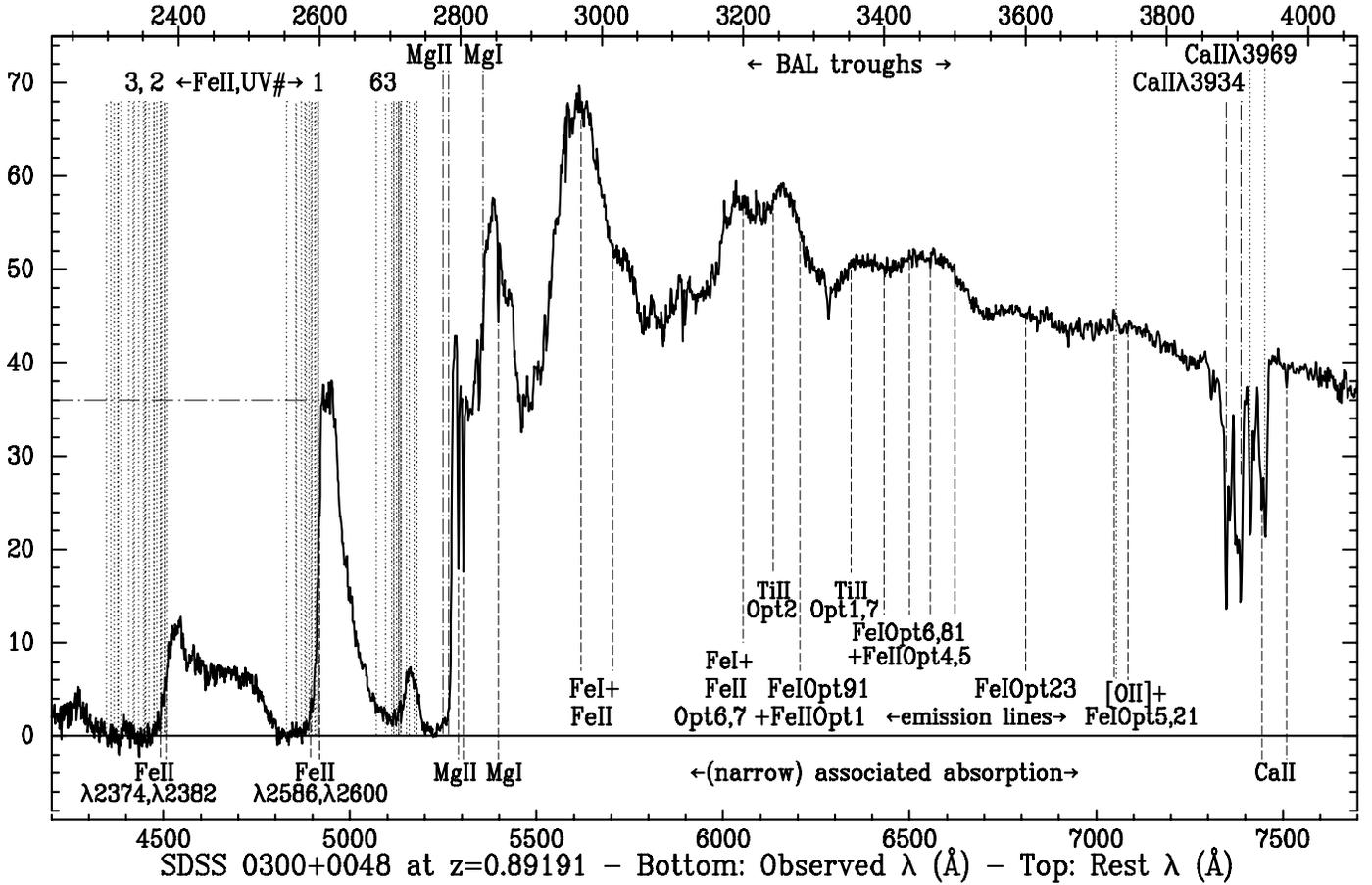}
\caption[]{ \singlespace 
A portion of the SDSS spectrum ($R \sim 1800$)
of SDSS J0300+0048 ($F_{\lambda}$ in units of
$10^{-17}$\,ergs cm$^{-2}$\,s$^{-1}$\,\AA$^{-1}$ vs. $\lambda$).
Selected broad absorption line troughs are labelled along the top,
and the narrow associated absorption lines along the bottom,
to help place the UVES spectrum in context.
The likely emission lines are labelled just above the labels 
for associated absorption.
The horizontal dot-dashed line shows the continuum level used to normalize
the UVES spectrum shortward of 2600\,\AA\ rest-frame.
For a more comprehensive plot of the SDSS spectrum of this object, see H02.
}\label{f_0300sb2}
\end{figure}

\begin{figure}
\epsscale{1.7}
\plotone{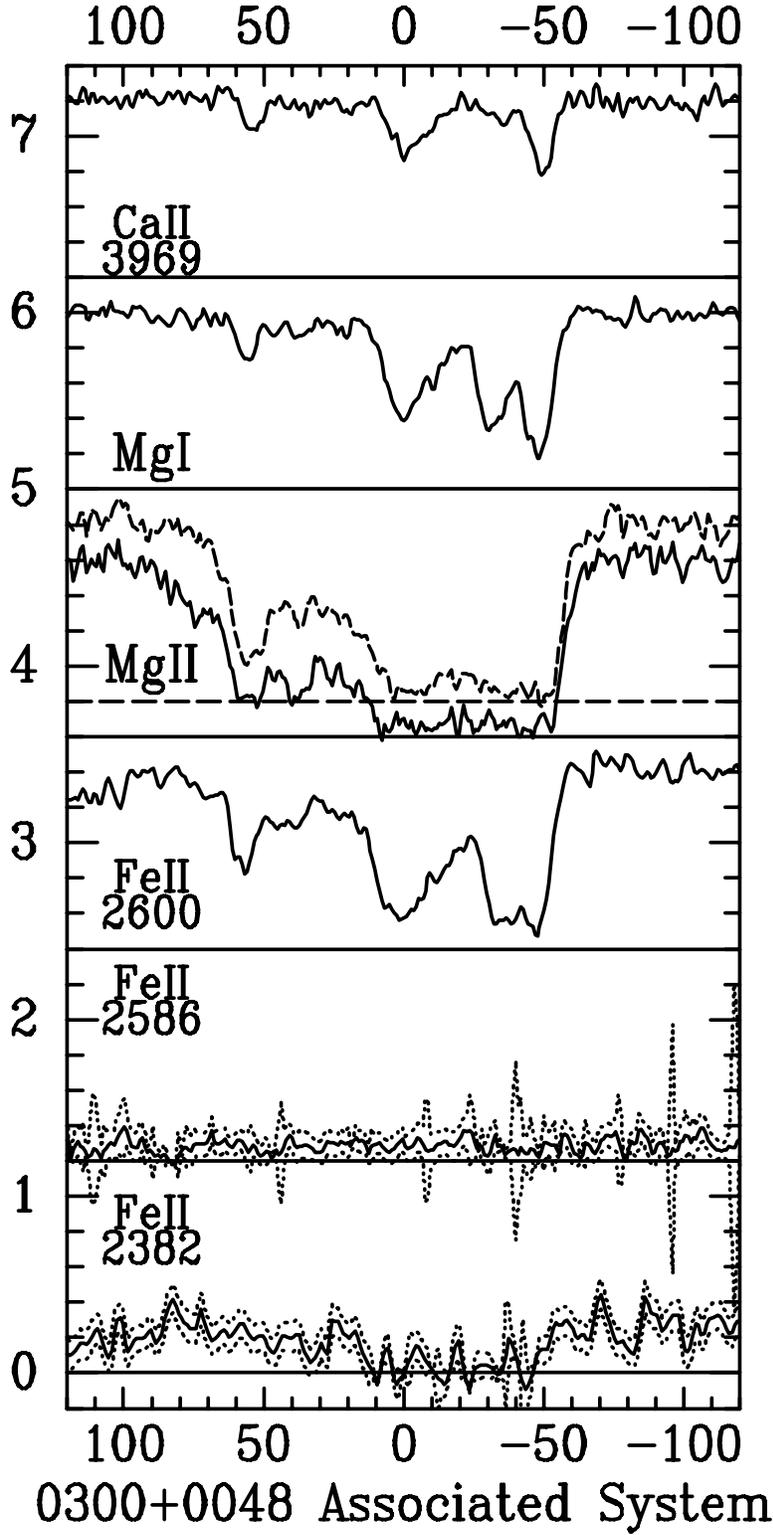}
\caption[]{ \singlespace 
Transitions in the associated absorption system in SDSS J0300+0048.
The vertical scale is the normalized flux density $F_{\lambda}$ and
the horizontal scale is the velocity in \kms\ (positive values for outflow).
Each transition shown has been offset from
the next by 1.2 normalized flux units in the vertical direction.  
Both lines of the \mgii\ doublet are plotted; 
the $\lambda$2803 line is plotted as a dotted line offset by +0.2 flux units.
For two of the \feii\ transitions dotted lines are plotted to
show the $\pm1\sigma$ uncertainty ranges.
}\label{f_6way}
\end{figure}

The key insight comes from 
considering the associated absorption in \feii\,$\lambda\lambda$2374,2382
and \feii\,$\lambda\lambda$2586,2600 --- lines that all 
arise from the lowest fine-structure level in the ground term of \feii.
Between the redshifted wavelengths of the associated \feii\,$\lambda$2600 and
\feii\,$\lambda$2586, the \feii\ UV1 BAL (absorption from ground and excited
states) removes most of the continuum, but not quite all (Fig. \ref{f_0300sb2}).
If the associated absorption arises far out in the quasar host galaxy,
we might expect to 
see \feii\,$\lambda$2586 absorb $\sim$93\% of the remaining continuum,
just as \feii\,$\lambda$2600 does.
The normalized flux and $\pm$1$\sigma$ uncertainties in this spectral
region are shown in the second panel from the bottom of Figure \ref{f_6way}.
Although the noise is large, there
is no evidence for the expected absorption even when the spectrum is binned.
However, there is clear evidence for \feii\,$\lambda$2382 
absorption 
(Figure \ref{f_6way}, bottom panel).
This spectral region is equally noisy, but in this case the absorption
{\em is} consistent with being 93\% of the surrounding
continuum.\footnote{The local continuum used to normalize the
\feii\,$\lambda$2586 region in Figure \ref{f_6way} is identical to that used for
\feii\,$\lambda$2600: it is flat in $F_{\lambda}$ at the level of the spectrum
just longward of \feii\,$\lambda$2600.  This same continuum was used to
normalize the \feii\,$\lambda$2382 region, for lack of a better assumption.
However, the normalization does not affect the interpretation of why
associated absorption from \feii\,$\lambda$2586 was not detected.} 
The same is true for \feii\,$\lambda$2374, 
while the \feii\,$\lambda$2344 region is too noisy for analysis.

\begin{figure} \epsscale{0.54} 
\plotone{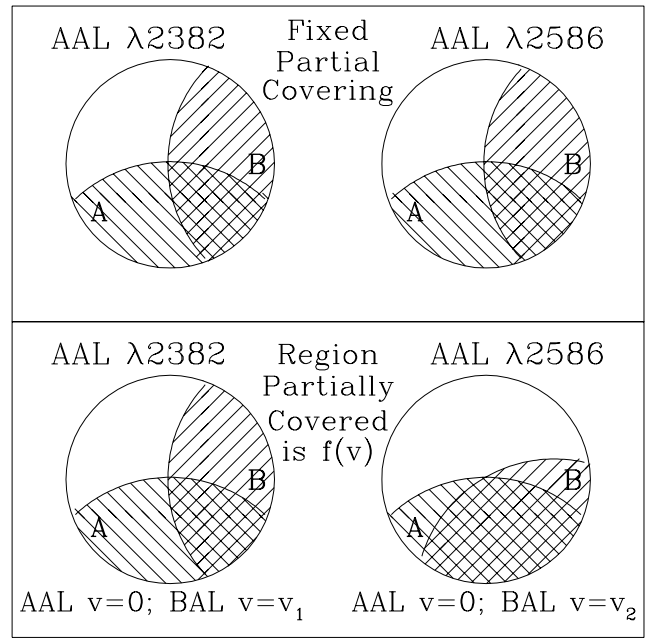}
\caption[]{ \singlespace 
Schematic model of a quasar (circle) and absorption with partial covering.
The lines slanting down and to the right show the source region
covered by the narrow, associated, ground-state \feii\ absorption (A), 
while the lines slanting up and to the right 
show the source region covered by the broad, excited \feii\ absorption (B).
The partial covering is shown at the redshifted wavelength of the associated
\feii\,$\lambda$2382\,\AA\ absorption on the left, and at the redshifted
wavelength of the associated \feii\,$\lambda$2586\,\AA\ absorption on the right.
As discussed in the text, the spatially fixed partial covering shown in the
top panel cannot explain our observations.  Instead, our data indicate that
the broad absorption trough covers {\em different} regions of the quasar
as a function of velocity, as shown in the bottom panel.
}\label{f_partcover0300}
\end{figure}

The apparent lack of \feii\,$\lambda$2586 absorption
rules out the simple picture of a BAL trough 
and an associated absorption line system (AAL) that each absorb 
spatially fixed regions of quasar emission,
shown schematically in the top panel of Figure \ref{f_partcover0300}.
This conclusion applies even if the BAL or AAL regions are not
well characterized by a single $C$ and $\tau$ \markcite{dka02}({de Kool} {et~al.} 2002b).
To explain our observations requires some variation with velocity 
in the quasar emission region covered by the BAL trough. 
The bottom panel of Figure \ref{f_partcover0300} is a sketch of
this spatially distinct, velocity-dependent partial covering. 

The reason the BAL region can partially cover spatially distinct emission
regions at the wavelengths of the two associated \feii\ transitions is that the
broad \feii\ absorption includes absorption from excited fine-structure levels,
while the narrow \feii\ absorption arises from only the ground level.
At the zero-velocity wavelength of associated \feii\,$\lambda$2382, 
the broad \feii\ absorption can be approximated as having some velocity $v_1$.
Because it consists of absorption in different \feii\ multiplet UV2
transitions (both ground and excited) at different velocities, 
$v_1 = {{1}\over{N}} \sum_i^N w_i v_i$, where $w_i$ is the
relative strength of line $i$ at velocity $v_i$.
Similarly, at the zero-velocity wavelength of associated
\feii\,$\lambda$2586, 
the broad \feii\ absorption will consist of ground and excited \feii\ 
multiplet UV1 transitions and can be approximated as having velocity 
$v_2 = {{1}\over{M}}\sum_j^M w_j v_j$.  
The velocities $v_1$ and $v_2$ will be similar {\em but not identical}.
Therefore, the covering factors can differ as well, both numerically
($C(v_1) \neq C(v_2)$) and in the spatial regions covered.
In SDSS J0300+0048, 
those different covering factors as a function of velocity allow
the associated and broad \feii\ absorbers to overlap the same spatial region
at the wavelength of associated \feii\,$\lambda$2586 but not at the wavelength
of associated \feii\,$\lambda$2382.  This wavelength difference translates
to a velocity difference $v_1 - v_2 \sim 1700$~\kms\ over which there
must be considerable variation in the regions being partially covered.
Similar effects have been seen before in FBQS 1408+3054 (\S\,6.1.2 of H02)
and NGC 3783 \markcite{gea03}({Gabel} {et~al.} 2003).

We cannot prove that the associated absorber is external to the BAL region,
but we have shown that such a geometry, coupled with the spatially distinct,
velocity-dependent partial covering that must be present, can explain our
observations.  We therefore used this `external' assumption 
to remove the associated \caii\ absorption from the broad \caii\ trough.

\end{appendix}

\begin{footnotesize}


\end{footnotesize}

\begin{deluxetable}{ccccccc}
\tablecaption{Column Densities
in the SDSS J0300+0048 BAL Outflow\label{t_colden}}
\tablehead{
\colhead{Velocity Range:} & 
\colhead{1697-3740~km\,s$^{-1}$} & \colhead{3740-5657~km\,s$^{-1}$} & \colhead{5657-8221~km\,s$^{-1}$} & \colhead{Total}
}
\startdata
\ion{Fe}{2}* & $\leq (0.04\pm0.16) \times 10^{15}$ & $\geq (1.34\pm0.09) \times 10^{15}$ & $\geq (2.95\pm0.58) \times 10^{15}$ & $\geq (4.29\pm0.59) \times 10^{15}$ \\
\ion{Mg}{2} & $\geq (1.90\pm0.34) \times 10^{15}$ & $\geq (2.51\pm0.09) \times 10^{15}$ & $\geq (2.20\pm0.07) \times 10^{15}$ & $\geq (6.61\pm0.36) \times 10^{15}$ \\
\ion{Mg}{1} & $\geq (1.54\pm0.02) \times 10^{13}$\tablenotemark{a} & $\geq (2.84\pm0.02) \times 10^{13}$ & \nodata & $\geq (4.38\pm0.03) \times 10^{13}$\tablenotemark{a} \\
\ion{He}{1}* & $\leq 2.5           \times 10^{14}$ & $\leq 2.3           \times 10^{14}$ & $\leq 3.1         \times 10^{14}$ & $\leq 7.9 \times 10^{14}$ \\
\ion{Ca}{2} & $     (5.22\pm0.75) \times 10^{14}$ & $     (1.91\pm0.87) \times 10^{14}$ & $\leq (0.3\pm1.2) \times 10^{12}$ & $(7.13\pm1.15) \times 10^{14}$ \\
\ion{Ca}{1} & $\leq 0.2            \times 10^{12}$ & $\leq 0.8           \times 10^{12}$ & $\leq 0.7         \times 10^{12}$ & $\leq 1.7 \times 10^{12}$ \\
\enddata
\tablenotetext{a}{This \ion{Mg}{1} column density is a conservative lower limit;
more liberal assumptions yield almost exactly the same column density as at
3740-5657~km\,s$^{-1}$.  In that case, the total \ion{Mg}{1} column would be
$(5.66\pm0.03) \times 10^{13}$.}
\tablecomments{All column densities are given in units of cm$^{-2}$.
The asterisks in the first column denote ions whose column densities
are measured in excited states rather than the ground state.}
\end{deluxetable}


\end{document}